\documentclass[%
 aip,
 amsmath,amssymb,
 reprint,%
]{revtex4-1}

\usepackage{graphicx}
\usepackage{dcolumn}
\usepackage{bm}
\usepackage[utf8]{inputenc}
\usepackage[T1]{fontenc}
\usepackage{mathptmx}
\usepackage{etoolbox}
\usepackage{array}

\makeatletter
\def\@email#1#2{%
 \endgroup
 \patchcmd{\titleblock@produce}
  {\frontmatter@RRAPformat}
  {\frontmatter@RRAPformat{\produce@RRAP{*#1\href{mailto:#2}{#2}}}\frontmatter@RRAPformat}
  {}{}
}%
\makeatother
\begin{document}

\preprint{AIP/123-QED}

\title{Two-dimensional electronic spectroscopy from first principles}

\newcommand{\UOL}{Carl von Ossietzky Universit\"at Oldenburg, Physics Department and Center for Nanoscale Dynamics (CeNaD), D-26129 Oldenburg, Germany}
\newcommand{\HU}{Humboldt-Universit\"at zu Berlin, Physics Department and IRIS Adlershof, D-12489 Berlin, Germany}

\author{Jannis Krumland}
\affiliation{\UOL{}}
\affiliation{\HU{}}
\email{jannis.krumland@physik.hu-berlin.de}

\author{Michele Guerrini}
\affiliation{\UOL{}}

\author{Antonietta De Sio}
\affiliation{\UOL{}}

\author{Christoph Lienau}
\affiliation{\UOL{}}

\author{Caterina Cocchi}
\affiliation{\UOL{}}
\affiliation{\HU{}}
\email{caterina.cocchi@uni-oldenburg.de}

\date{\today}

\begin{abstract}
The recent development of multidimensional ultrafast spectroscopy techniques calls for the introduction of computational schemes that allow for the simulation of such experiments and the interpretation of the corresponding results from a microscopic point of view.
In this work, we present a general and efficient first-principles scheme to compute two-dimensional electronic spectroscopy maps based on real-time time-dependent density-functional theory.
The interface of this approach with the Ehrenfest scheme for molecular dynamics enables the inclusion of vibronic effects in the calculations based on a classical treatment of the nuclei.
The computational complexity of the simulations is reduced by the application of numerical advances such as branching techniques, undersampling, and a novel reduced phase cycling scheme, applicable for systems with inversion symmetry.
We demonstrate the effectiveness of this method by applying it to prototypical molecules such as benzene, pyridine, and pyrene. 
We discuss the role of the approximations that inevitably enter the adopted theoretical framework and set the stage for further extensions of the proposed method to more realistic systems.
\end{abstract}

\maketitle

\section{Introduction}

The most recent advances in ultrafast spectroscopies have unlocked unprecedented opportunities to gain insight into the fundamental properties of materials and their dynamics by following the response of the system on the natural spatiotemporal scale of its constituents, such as electrons, holes, phonons, etc.
This way, it is possible not only to access the behavior of individual excitations but, more relevantly, their interplays.
Multidimensional spectroscopies are the most powerful techniques to achieve an understanding of the (coherent) couplings between different degrees of freedom in an excited system.\cite{wrig11arpc,cund-muka13pt,ruet+14pnas}
Among them, two-dimensional electron spectroscopy (2DES; all abbreviations are listed in Appendix~\ref{sec:abbreviations}) has established itself as the state-of-the-art approach for accessing the structure and dynamics of materials beyond the linear regime.\cite{jona2003arpc,Cho2006,zigm+22cpr,bisw+2022cr,Brixner+2005nat}
The output of 2DES consists of maxima identified by two time or frequency axes, namely, excitation and detection,\cite{Gelzinis_2019,coll21jpcc} unveiling the contributions to the excitation from the various degrees of freedom (electronic, vibrational, \textit{etc.}) and their couplings;\cite{desi-lien17pccp,zueh+21arpc} it also provides quantitative indications about (de)excitation pathways, including conical intersections.\cite{desi+21natn,till+21jpcl,cho+22jacs}
Due to the wealth of information encoded in 2DES, the interpretation of its spectra is far from trivial.\cite{Duan+2017pnas}

The most common approach to compute 2DES spectra is the density matrix (DM) formalism,~\cite{hamm_zanni_2011,Mukamel1995,geli+09acr,wehn+14jcp,troi23jcp}, a scheme based on the summation of the Feynman diagrams associated with the specific pathways of the scattered beam. This method is analytical and intuitive and has proven itself particularly useful to interpret the features of 2DES in complex materials such as halide perovskites,\cite{nguy+19jpcl,nguy+23natcom,Monahan+2017jpcl} organic semiconductors,\cite{desi+16natcom,desi+2019zna,desi+21natn,Scholes+2011nc,Collini+2009sci,Bakulin+2016nc} and photosynthetic biosystems.\cite{Collini+2010nat,Engel+2007nat,rivalta2014}
Yet, the DM approach does not work \textit{ab initio}, \textit{i.e.}, it needs empirical parameters to construct system-dependent Hamiltonians.
This is very effective for direct comparisons with experimental data but limits the applicability of the method and its predicting power to systems that have been already measured.
The substitution of empirical parameters with those extracted from quantum-chemical approaches~\cite{rivalta2014,sega-mart18tcc} represents a reasonable way to circumvent this issue. 
However, the availability of a full-fledged first-principles approach for 2DES would offer unprecedented opportunities to explore and predict the properties of materials with the tools provided by this technique.
Furthermore, non-perturbative real-time approaches might offer superior scaling with respect to the system size, compared to methods based on the perturbative construction of response functions.~\cite{geli+22cr}

The current trends in computational materials science of embedding \textit{ab initio} calculations into automatized workflows that guarantee data provenance through their database storage and integrated post-processing tool for data analysis~\cite{Kulik_2022} call for routines that are integrated into a single package. 
Density-functional theory (DFT) and its real-time, time-dependent extension (RT-TDDFT) are particularly promising methods for undertaking this task. 
Not only do they enable the simulation of (in principle) any material regardless of its composition and dimensionality, but they have also proven to offer an optimal trade-off between accuracy and numerical efficiency in the simulation of transient optical spectra of numerous systems, ranging from bulk crystals~\cite{otob+08prb,sato+14prb,wach+14prl,tanc+17natcom,liu+20cms} to two-dimensional (2D) materials,\cite{tanc-rubi18sa,hash+22apl} organic molecules,\cite{lopa-govi11jctc,cocc+14prl,krum+20jcp,guan+21pccp} and interfaces.\cite{rozz+13natcom,falk+14sci,urat-naka23jpcl}

Motivated by these successes, herewith we present an efficient computational scheme to calculate 2DES spectra from first principles using RT-TDDFT. This approach offers several advantages: (i) being \textit{ab-initio}, it is general and can be used for different systems as it does not rely on model Hamiltonians and empirical parameters; 
(ii) it works in real-time and it is intrinsically non-perturbative. This means that all the Feynman pathways, including all inter-pulse time orderings, are automatically included at once;
(iii) there is no limitation on the pulse amplitude, \textit{i.e.}, any nonlinear order response can be computed, in principle. This last point in particular can be exploited to explore strong nonlinear ultrafast dynamics.

After discussing the theoretical foundation of this method,
we present three approaches to tame the computational complexity associated with corresponding simulations: branching, undersampling, and a reduced phase cycling scheme that can be employed for inversion-symmetric systems. Establishing an interface with the open-source package \textsc{Octopus},~\cite{Tancogne_et_al_2020} we simulate the photo-induced dynamics of the prototypical molecules benzene, pyridine, and pyrene.
We inspect the response of these compounds, characterized by a similar backbone but by different symmetries, in the linear and nonlinear regime, 
obtaining results that are consistent with the energies and symmetries of the many-body states within the adopted approximations for the exchange-correlation functional.
Taking advantage of the built-in interface between TDDFT and the Ehrenfest scheme for molecular dynamics available in \textsc{Octopus},\cite{marq+03cpc} we additionally monitor in the 2DES spectra the coupled electron-nuclear response of the systems.
In this discussion, we point out the advantages of the proposed approach as well as the current limitations mainly based on the underlying approximations for the exchange-correlation functional and the nuclear motion, which will be improved in future work.

This paper is structured as follows.
In Sec.~\ref{sec:2DES}, we introduce the general principles of 2DES (Sec.~\ref{sec:2DES-general}) to prepare the ground for presenting the underlying theory of the proposed formalism developed in the framework of RT-TDDFT (Sec.~\ref{sec:methods}). The calculation workflow is provided in Sec.~\ref{sec:implementation}. In Sec.~\ref{sec:results}, the introduced method is applied to benzene, pyridine, and pyrene. For the latter, the analysis is extended to inspect vibronic effects.
The summary and the conclusions of this work are in Section~\ref{sec:conclusions}.

\section{Two-dimensional spectroscopy}
\label{sec:2DES}

\subsection{General Principles}
\label{sec:2DES-general}

In 2DES, the measured signal is governed by the third-order polarization induced in the sample by the interaction with a sequence of three ultrashort optical pulses as a function of the optical frequency and of two experimentally controlled time delays: the coherence time $\tau$ between the first two pump pulses and the waiting time $T$ between the pump pulse and the third (probe) pulse. The pulse sequence is schematically sketched in Fig.~\ref{fig:pulse_train_and_feynman_diagrams}a). 
The nonlinear third-order signal radiated from the sample emerges after its interaction with the probe pulse and it is measured during the detection time $t$. For each waiting time $T$, 2D energy-energy correlation maps are obtained by Fourier transform of the measured signal with respect to $\tau$ and $t$ time delays. The third-order polarization in the time domain is determined by the third-order optical response of the sample convoluted with the optical pulses. The nonlinear response function (\textit{i.e.}, the nonlinear optical susceptibility) describes the microscopic behavior of the material, \textit{i.e.}, it describes the free evolution and the details of the light-matter interaction of the sample. It is thus the quantity of interest to assess the response of the examined material. 
Depending on the time ordering of the pulse sequence and the phase matching dictated by the experimental configuration, different quantum pathways contributing to the third-order response can be probed. These contributions can be generally separated into rephasing and non-rephasing signals, and their sum defines the absorptive spectra that we discuss in the following.\cite{hamm_zanni_2011}

In the most general experimental configuration of 2DES, the three optical pulses propagate non-collinearly and impinge on the sample under small angles, thus with different wavevector orientations. In this case, the nonlinear signal from the sample is emitted in the direction that satisfies momentum conservation under a fourth wavevector.\cite{jona2003arpc} In a 2DES experiment, the optical pulses are usually resonant with (some of) the optical transitions of the sample to investigate. The first optical pulse induces a coherence between the ground and the excited states of the material. This induces a coherent polarization that evolves during the coherence time until the arrival of the second pulse. The latter interacts with the polarization and generates a spatial population grating in the sample, which further evolves during the waiting time.\cite{bisw+2022cr}
Finally, the third pulse scatters off the population grating in the phase-matched direction defined by the experimental configuration. The nonlinear field re-emitted by the sample, induced by the sequence of optical pulses, can be retrieved by spectral interferometry with a local oscillator, which does not interact with the sample.\cite{bris+2008oe, bris+2009rsi}

In the partially collinear configuration scheme, collinearly propagating pump pulses are used. As such, in setup, the 2DES experiment can be thought of as an extension of a more conventional pump-probe one. The difference is that, instead of a single pump pulse, a pair of collinearly propagating phase-locked and time-delayed pump pulses are used in the excitation arm of the setup.\cite{desi+2016natcom} The phase-locked pump pulse pair gives rise to an interference spectrum in the frequency domain whose fringe spacing is inversely proportional to the coherence time.\cite{desi-lien17pccp, desi+2019zna} This Fourier-transform approach in the excitation allows for overcoming the trade-off of maintaining short pulse durations and selective spectral excitation of specific samples' resonances. The excitation by the pump-pulse pair allows for obtaining a frequency-resolved excitation axis and thus 2D maps. The detection axis is determined by the spectral range of the probe pulse. After the interaction with the sample, the pump-induced changes in the transmitted or reflected spectrum of the probe pulse are recorded with a spectrometer.\cite{desi+2016natcom, nguy+23natcom} In this implementation, the nonlinear signal is emitted collinearly with the probe pulse and thus the probe itself is used to retrieve the nonlinear signal without the need for an external local oscillator. Because rephasing and non-rephasing pathways are emitted collinearly in this configuration, purely absorptive spectra are readily available, whereas the single rephasing and non-rephasing signals can be separated, if required, by implementing for example phase cycling schemes.\cite{myer+08oe,cho+2018jpcl}

\subsection{Theory}\label{sec:methods}
In this section, the concepts laid out in Sec.~\ref{sec:2DES-general} are formalized, arriving at equations that are fit to be solved numerically. The electronic structure theory of choice, (RT-TD)DFT, is briefly discussed.
\subsubsection{From Nonlinear Polarization to 2DES}

\begin{figure}
    \centering
    \includegraphics[width=0.47\textwidth]{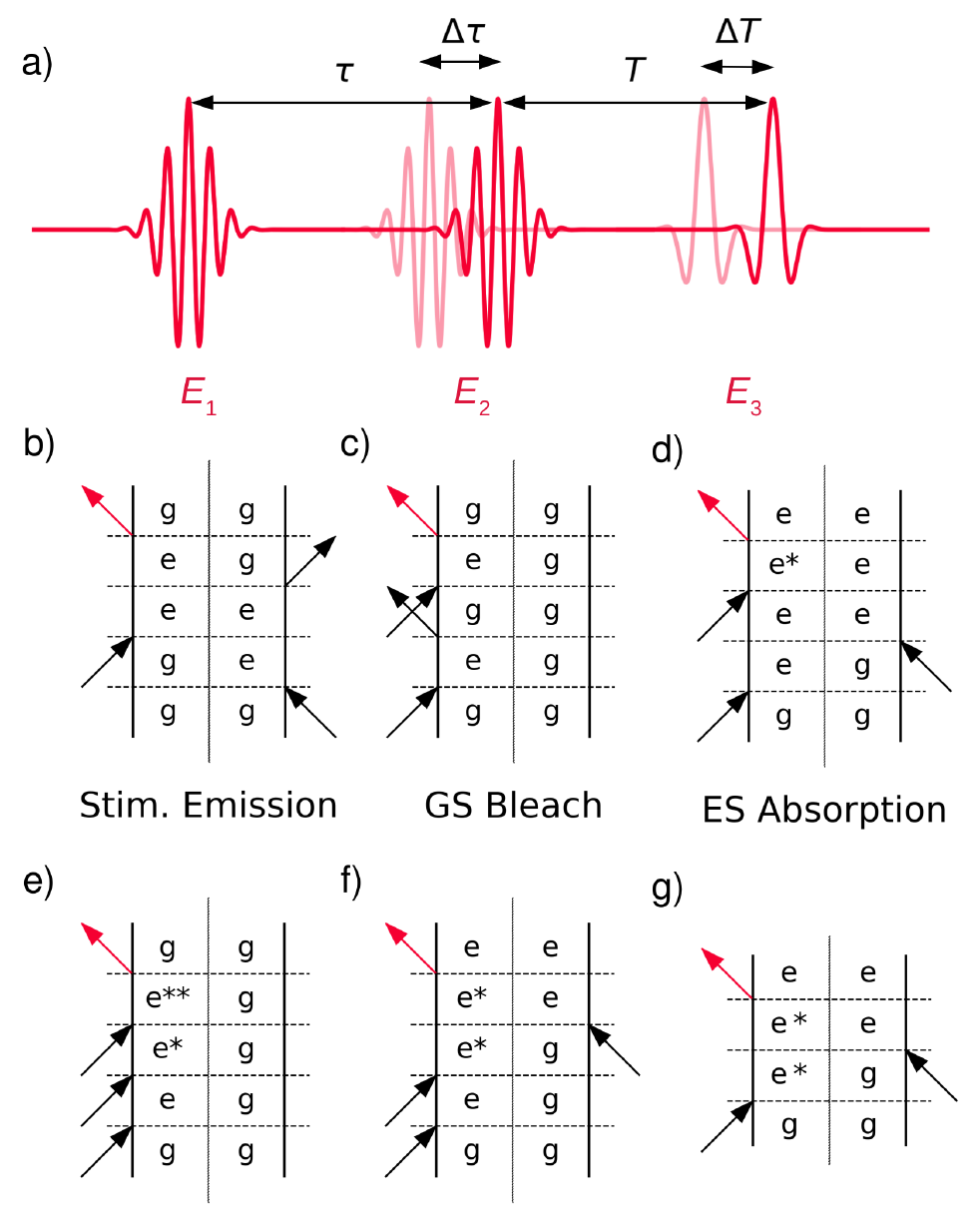}
    \caption{a) Sketch of a pulse train with coherence time $\tau$ and waiting time $T$; $\Delta\tau$ and $\Delta T$ are the corresponding sampling periods. b)-d) Selected double-sided Feynman diagram representing processes pertaining to the third-order dipole $\mu^{(3)}_{\pm\textbf{k}_3}$ that emits a field interfering with the probe pulse. e)-g) Selected processes that contribute to the total dipole moment $\mu$ induced by the pulse train, but do not belong to $\mu^{(3)}_{\pm\textbf{k}_3}$. The process in panel g) represents a second-order process that is only possible in a system without inversion symmetry. In all diagrams, $\mathrm{g}$ stands for the ground state, while $\mathrm{e}$, $\mathrm{e}^*$, and $\mathrm{e}^{**}$ denote excited states.}
    \label{fig:pulse_train_and_feynman_diagrams}
\end{figure}

We consider a large number of non-interacting molecules forming a macroscopic sample, \textit{e.g.}, a film or a solution. The photoexcitation of this system gives rise to dipolar oscillations within the constituting units. Such a collection of coherently oscillating dipoles $\mu$ with local number density $n$ itself emits an electric field that, at large distances, is proportional to the inverse of the time derivative of the induced local polarization, $P=n\mu$.~\cite{feyn+1965flp, murray1978laser} We consider specifically the field $E^{(3)}$ resulting from the respective third-order components $P^{(3)}$ and $\mu^{(3)}$,
\begin{align}\label{eq:p3_to_sig}
E^{(3)}\propto-\frac{\text dP^{(3)}}{\text dt} = -n\frac{\text d\mu^{(3)}}{\text dt}.
\end{align}
$P^{(3)}$ can, in turn, be calculated through a nonlinear convolution\cite{Mukamel1995}:
\begin{align}
\label{eq:third_order_dipole}
    P^{(3)}(t) = n&\mu^{(3)}(t) = \int_{-\infty}^{\infty} \mathrm dt_1\int_{-\infty}^{\infty}\mathrm dt_2 \int_{-\infty}^{\infty} \mathrm dt_3\, S^{(3)}(t_1,t_2,t_3) \times \nonumber\\ 
    \times & E(t-t_3) E(t-t_3-t_2) E(t-t_3-t_2-t_1),
\end{align}
where we assume an external electric field $E$ that is composed of three pulses,
\begin{equation}
\label{eq:pulse_seq}
    E(t)= E_1(t + T + \tau) + E_2(t + T) + E_3(t),
\end{equation}
\textit{i.e.}, two equivalent phase-locked pump pulses $E_1 = E_2$, as well as the final probe pulse $E_3$ [Fig.~\ref{fig:pulse_train_and_feynman_diagrams}a)].  
The third-order response function in Eq.~\eqref{eq:third_order_dipole} can be written as~\cite{Mukamel1995} 
\begin{widetext}
\begin{align}
\label{eq:third_order_response_function}
S^{(3)}(t_1,t_2,t_3) \propto \left( \frac{i}{\hbar} \right)^3  \theta(t_1)\theta(t_2)\theta(t_3) \mathrm{Tr}\left\lbrace\hat\mu(t_1+t_2+t_3) \left[\hat\mu(t_1+t_2), \left[\hat\mu(t_1),\left[\hat\mu(0), \hat\rho_\mathrm{GS} \right] \right] \right]\right\rbrace
\end{align}
\end{widetext}
where $\theta$ is the Heaviside function enforcing causality in each time domain, $\hat\mu(t)$ is the dipole operator in the interaction picture, and $\hat\rho_\mathrm{GS}$ is its initial ground-state (GS) DM. For simplicity, we ignore here the vectorial nature of $E$, $P$, and $\mu$, which would turn $S^{(3)}$ into a fourth-rank tensor. In the nonlinear regime, the polarization from the collectively oscillating microscopic dipoles emits electric fields in multiple directions that can be written as linear combinations of the wave vectors of the incident fields,
\begin{align}
    n_1\textbf{k}_1 + n_2\textbf{k}_2 + n_3\textbf{k}_3,
\end{align}
where $\textbf{k}_j$ is the wave vector associated with the field $E_j$. We are specifically interested in the component $\mu_{\pm\textbf{k}_3}^{(3)}$ of $\mu^{(3)}$ that is responsible for emission in the direction of $\textbf{k}_3$, \textit{i.e.}, along the propagation direction of the probe pulse. It is this component that determines the absorptive 2DES, $A_\mathrm{2D} = A_\mathrm{2D}(\omega_\mathrm{exc},T,\omega_\mathrm{det})$. We write
\begin{align}
\mu^{(3)}_{\pm\textbf{k}_3} = \mu^{(3)}_{\pm\textbf{k}_3}(\tau, T, t),
\end{align}
where $\tau$ and $T$ are the controlled delays between the incident electric fields in Eq.~\eqref{eq:pulse_seq}. To map out this three-variable function, several calculations with varying delays $\tau$ and $T$ between the three pulses are to be carried out. According to Eq.~\eqref{eq:p3_to_sig}, the signal field picked up by the detector in the probe direction is
\begin{align}
    E_{\pm\textbf{k}_3}^{(3)}(\tau, T, t) \propto -\frac{\text{d} \mu_{\pm\textbf{k}_3}^{(3)}}{\text{d} t}.
\end{align}
The Fourier transform of this field with respect to $\tau$ ($\rightarrow\omega_\mathrm{exc}$) and $t$ ($\rightarrow\omega_\mathrm{det}$) gives

\begin{align}
    E_{\pm\textbf{k}_3}^{(3)}(\omega_\mathrm{exc}, T, \omega_\mathrm{det}) \propto i\omega_\mathrm{det}\mu_{\pm\textbf{k}_3}^{(3)}(\omega_\mathrm{exc}, T, \omega_\mathrm{det}),
\end{align}
while the waiting time $T$ is usually left untransformed. The 2DES is finally calculated as
\begin{align}
    A_\mathrm{2D}(&\omega_\mathrm{exc}, T, \omega_\mathrm{det}) \nonumber\\
    &= \Re\left\{\sum_\pm E_{\pm\textbf{k}_3}^{(3)}(\pm\omega_\mathrm{exc}, T, \omega_\mathrm{det})/E_3(\omega_\mathrm{det})\right\} \nonumber\\
    &\propto -\omega_\mathrm{det}\Im\left\{\sum_\pm \mu_{\pm\textbf{k}_3}^{(3)}(\pm\omega_\mathrm{exc}, T, \omega_\mathrm{det})/E_3(\omega_\mathrm{det})\right\}.
    \label{eq:2des_from_dipole}
\end{align}
The normalization with respect to $E_3$ removes the dependence on the spectral shape of the probe field along the $\omega_\mathrm{det}$ axis. Mirroring spectral features at frequencies $\omega_\mathrm{exc}<0$ into the positive region $\omega_\mathrm{exc}>0$ through the $\pm\omega_\mathrm{exc}$-sum corresponds to superposing rephasing and non-rephasing spectra. This step yields the \emph{purely absorptive} spectrum, which is known to have optimal frequency resolution.~\cite{hamm_zanni_2011, Mukamel1995}

For the analysis of the spectra, it is helpful to visualize the occurring processes in terms of \textit{double-sided Feynman diagrams}, which depict the time evolution of the DM~\cite{hamm_zanni_2011} [Fig.~\ref{fig:pulse_train_and_feynman_diagrams}b)-g)]. The nonlinear response function in Eq.~\eqref{eq:third_order_response_function} can be decomposed as
\begin{align}\label{eq:third_order_response_resolution}
S^{(3)}(t_1,t_2,t_3) = \sum_{n}S^{(3)}_n(t_1,t_2,t_3),
\end{align}
where each $S^{(3)}_n$ is a different third-order process or pathway composed of three instantaneous field-system interactions [black arrows in Fig.~\ref{fig:pulse_train_and_feynman_diagrams}b)-g)] and the final dipole measurement at the detector [red arrows in Fig.~\ref{fig:pulse_train_and_feynman_diagrams}b)-g)].
Features visible in the $\textbf{k}_3$ direction and thus manifesting themselves in $A_\mathrm{2D}$ correspond to stimulated emission (SE), ground-state bleach (GSB), and excited-state absorption (ESA) processes [Fig.~\ref{fig:pulse_train_and_feynman_diagrams}b)-d)]. Third-order processes emitting in other directions [Fig.~\ref{fig:pulse_train_and_feynman_diagrams}e)-f)] do not give rise to visible features in $A_\mathrm{2D}$, and neither do second-order ones that are nonzero in systems without inversion symmetry  [Fig.~\ref{fig:pulse_train_and_feynman_diagrams}g)]. Since the output of our simulations is not the direction-resolved emitted field but the total dipole moment induced in a single molecule, it is necessary to post-process this quantity in order to filter out contributions that do not emit a field along the probe direction. This is accomplished by \textit{phase cycling}, which is described in the following.

\subsubsection{Isolating the dipole component emitting in the probe direction}
The computation of the 2D spectrum in Eq.~\eqref{eq:2des_from_dipole} requires the extraction of those contributions of the nonlinear polarization in Eq.~\eqref{eq:third_order_response_resolution} that emit a field along the probe beam direction ($\textbf{k}_3$). When the dipole moment is computed in a non-perturbative framework like in the present case, this can be accomplished by virtue of phase-cycling.~\cite{seid+1995jcp, meye+enge2000apb} The general idea of phase cycling is to suppress the undesired linear pump and probe field contributions, as well as the mutual interference terms in the second and third order, from the detected signal. For this, a phase modulation of the incident fields is introduced. 
\\
the total material polarization $P(\textbf{r},t)$ at position $\textbf{r}$ and time $t$ is decomposed into spatial Fourier components,~\cite{meye+enge2000apb}
\begin{align}\label{eq:fourier_decomposition_polarization}
    P(\textbf{r}, t) = \sum_{n_1, n_2, n_3}P_{n_1, n_2, n_3}(t) e^{i(n_1\textbf{k}_1 + n_2\textbf{k}_2 + n_3\textbf{k}_3)\cdot\textbf{r}}, 
\end{align}
where the component $P_{n_1, n_2, n_3}$ is responsible for emission along $n_1\textbf{k}_1 + n_2\textbf{k}_2 + n_3\textbf{k}_3$. In lowest order, the absolute values of the integer coefficients $n_j$ of the wave vector can be identified with the number of interactions between the system and the field $E_j$. For the pump-probe setup, we have $\textbf{k}_1=\textbf{k}_2$, and the desired result of the wave vector superposition is $\pm\textbf{k}_3$, corresponding to the component emitting in probe direction. Consequently, we obtain $n_1 + n_2 = 0$ and $n_3 = \pm 1$. To exclude the linear component ($n_1 = n_2 = 0$, $n_3 = 1$), we furthermore demand $|n_1| = |n_2| = 1$ to isolate the lowest nonlinear components. The superimposed terms in Eq.~\eqref{eq:fourier_decomposition_polarization} can be disentangled by evaluating $P$ (or, in practice, the dipole moment $\mu$ of a single molecule) with different phase offsets $\varphi_1$, $\varphi_2$, and $\varphi_3$ of the corresponding electric fields.\cite{meye+enge2000apb} The ($n_1$, $n_2$, $n_3$) component in Eq.~\eqref{eq:fourier_decomposition_polarization} thus gains a second phase factor,
\begin{align}\label{eq:phase_factor}
    e^{i(n_1\varphi_1 + n_2\varphi_2 + n_3\varphi_3)},
\end{align}
which can be exploited to suppress certain contributions and enhance others. As an example, the contribution with $n_1=2$, $n_2=0$, and $n_3 = 1$, which does not emit along $\textbf{k}_3$, can be removed by adding the $P$ resulting from calculations with $\varphi_1=0$ and $\varphi_1=\pi/2$, because the sum of the corresponding phase factors [Eq.~\eqref{eq:phase_factor}] vanishes as long as $\varphi_3$ is kept constant. Additional details about this procedure are reported in Appendix~\ref{sec:appendix_phase_cycling}, including the number of phases necessary to extract the desired contributions in systems with and without inversion symmetry.

\subsubsection{Solving Sampling-Related Problems}\label{sec:sampling_and_undersampling}

 At the current stage, the employed \textit{ab initio} method does not feature interactions with the environment beyond Ehrenfest nuclear dynamics. There is no mechanism that would lead to a loss of coherence, nor does the system experience population relaxation. The former implies that dipolar oscillations persist indefinitely, which is problematic from a computational point of view. It generally leads to discontinuities when cutting off discretized time series and Fourier-transforming them, as a discrete Fourier transform implicitly assumes a periodic continuation of the signal, inevitably causing artifacts. To mitigate this issue, the dipole moment is damped between the two pump pulses as well as after the probe pulse. A \textit{raised-cosine} damping function
\begin{align}\label{eq:damping_function}
    D(t) = \cos^2\left(\frac{\pi t}{2\tau_d}\right)
\end{align}
is applied during both the coherent and the detection times, \textit{i.e.},
\begin{align}\label{eq:third_order_dipole_triple_variable}
    \mu^{(3)}_{\pm\textbf{k}_3}(\tau, T, t) \rightarrow \mu^{(3)}_{\pm\textbf{k}_3}(\tau, T, t)D(\tau)D(t).
\end{align}
The reason for the choice of the damping function in Eq.~\eqref{eq:damping_function} is twofold: On the one hand, the initial decay is more gentle with respect to exponential damping, which is the usual function to model coherence decay. This has the advantage of requiring a shorter period of post-pulse propagation time to achieve the same spectral resolution when using longer exponential damping. On the other hand, the damping function Eq.~\eqref{eq:damping_function} suppresses the signal exactly (rather than asymptotically) to zero after the fixed \textit{dephasing time} $\tau_d$, which represents an upper limit for both $\tau$ and $t$. No damping is featured for the waiting time, during which the dynamics are qualitatively different. The dipole moment for the initial time along both $\tau$ and $t$ in the discretized time series is multiplied by a factor of 1/2, preventing further spectral artifacts.\cite{otti+1986jmr}

The dipole moment required to obtain 2DES is a function of three times [Eq.~\eqref{eq:third_order_dipole_triple_variable}]. It is recommended to sample the function along the $\tau$ and $T$ axes as coarsely as possible in order to reduce the computational complexity of the simulations. The pump pulses can excite within a certain spectral range, and the sampling frequency $2\pi/\Delta\tau$ of the coherence time needs to be at least twice as high as the upper edge of this window to avoid artifacts due to aliasing.\cite{nyquist_sampling} The number of sampling points along the $\tau$ axis amounts to $\tau_d/\Delta\tau$. This number can be quite large and represent a serious bottleneck. 
However, it is possible to reduce this number by \textit{undersampling} the signal along the $\tau$ dimension in a controlled and lossless fashion. Undersampling, \textit{i.e.}, the choice of a too-large sampling period, represents a violation of the Nyquist-Shannon sampling condition. This generally leads to artifacts in the form of aliasing, where the signal components larger than half the sampling frequency are mirrored back into the low-frequency region, mixing with actual low-frequency components. This results in a loss of information since the superposition cannot be disentangled. However, if the signal is known to have no components at lower frequencies, the high-frequency components are mapped into an empty spectral region. In this scenario, no information is lost, provided that it is known around which frequencies the signal should have been to begin with. These conditions are met for the 2DES along the $\omega_{exc}$ axis since the first pump pulse only excites states close to its carrier frequency. Thus, oscillations during the coherence time are confined to a small spectral region, leaving lower-frequency domains free to be used for lossless remapping through undersampling. During post-processing, a constant frequency offset is then added to the $\omega_{exc}$ axis, shifting the signal to the correct spectral region. Lowering the $\tau$ sampling rate through this technique can reduce significantly the amount of calculations to be performed.
The dynamics during $T$ are much slower since in this period the system resides either in a static excited state [Fig.~\ref{fig:pulse_train_and_feynman_diagrams}b)-d)] or in an excited-state coherence, which tends to have frequencies in the infrared. Consequently, the sampling period $\Delta T$ along $T$ can be chosen relatively coarse.

\begin{figure*}
        \centering
        \includegraphics[width=0.95\textwidth]{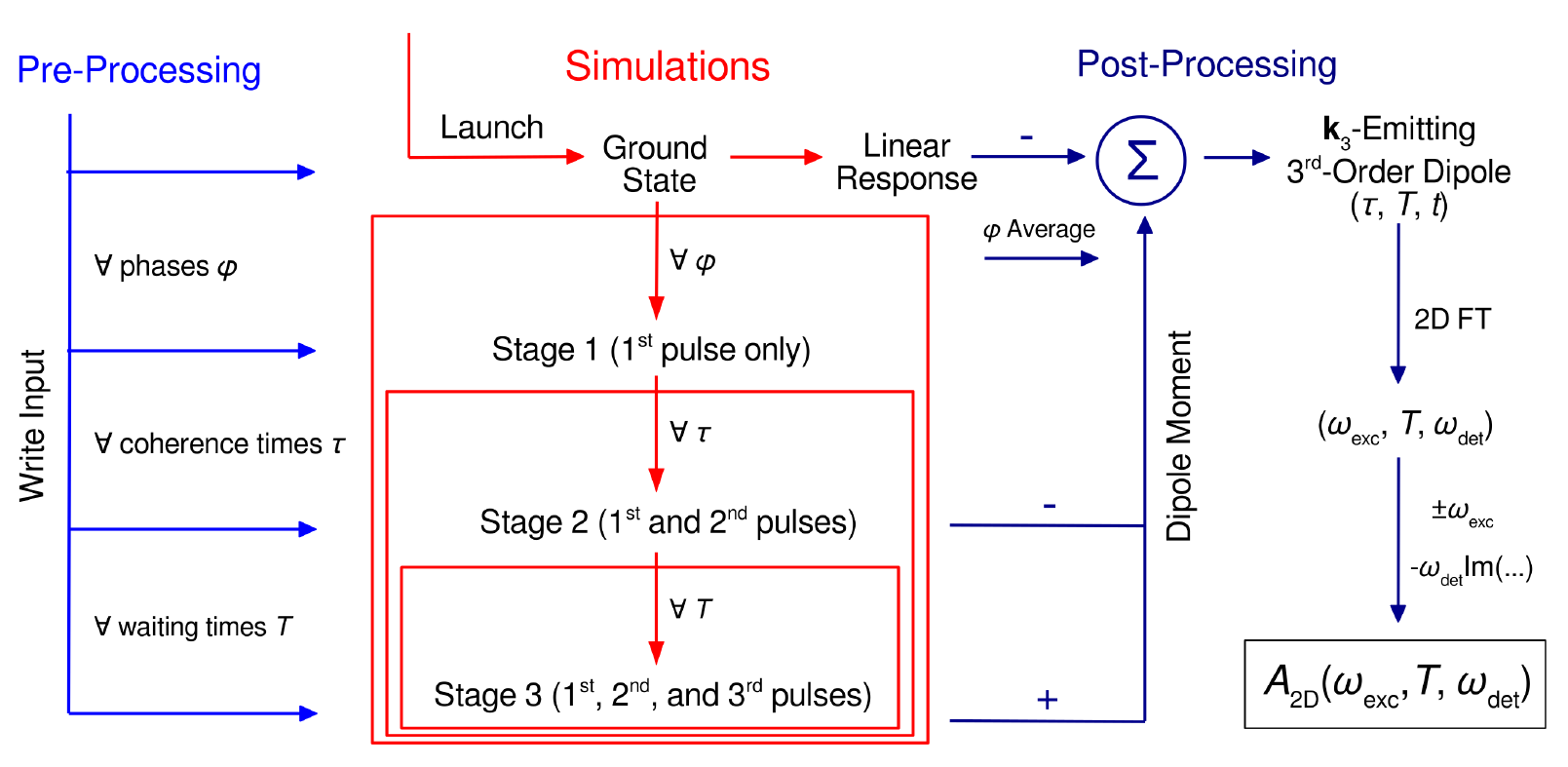}
        \caption{Computational workflow of the proposed 2DES simulation scheme. }
    \label{fig:workflow}
\end{figure*}

\subsubsection{RT-TDDFT Coupled with Ehrenfest Nuclear Dynamics}
The calculation of the dipole moment resulting from the application of a series of pulses is performed with RT-TDDFT. As a first step, the electronic structures of the molecules are calculated within DFT,\cite{hohenbergKohn1964pr, kohnSham1965pr} solving the Kohn-Sham (KS) equation,
\begin{align}\label{eq:ks_gs}
{\cal H}_\mathrm{GS}[\rho](R)\varphi(\mathbf{r}) = \varepsilon\varphi(\textbf{r}),
\end{align}
where $\varphi$ and $\varepsilon$ represent KS orbitals and eigenvalues, respectively; the KS Hamiltonian reads (we employ atomic units from hereon)
\begin{align}\label{eq:ks_hamiltonian}
    {\cal H}_\mathrm{GS}[\rho](R) = -\frac{\nabla^2}{2} + {V}_\mathrm{en}(\textbf{r}&|R)  + {V}_\mathrm{HXC}[\rho](\textbf{r}).
\end{align}
Here, ${V}_\mathrm{en}(\textbf{r}|R)$ is the attractive electrostatic potential at the electronic position $\textbf{r}$ due to the $N$ nuclei situated at positions $R=\{\textbf{R}_1,\ldots,\textbf{R}_N\}$:
\begin{align}
\label{Vext}
    {V}_\mathrm{en}(\textbf{r}&|R)= - \sum_{n=1}^N \frac{Z_n}{|\mathbf{r} - \mathbf{R}_n|},
\end{align}
where $Z_n$ is the charge of the $n$-th nucleus. ${V}_\mathrm{HXC}[\rho]$ is the Hartree-XC (HXC) potential
for a given electron density
\begin{align}\label{eq:density}
\rho(\mathbf{r}) = \sum_n^\mathrm{occ} |\varphi_n(\mathbf{r})|^2,
\end{align}
and contains the electrostatic repulsion between electrons as well as specifically quantum-mechanical aspects of the electron-electron interaction, which in practice must be approximated. The self-consistent solution of Eqs.~\eqref{eq:ks_gs} and \eqref{eq:density} yields the GS electron density, $\rho_\mathrm{GS}$, and the orbitals that constitute the starting point for the time evolution.

For TD calculations, the dynamical extension of Eq.~\eqref{eq:ks_gs} is adopted:
\begin{align}\label{eq:ks_td}
{\cal H}(t)\varphi(\mathbf{r}, t) = i\frac{\partial\varphi}{\partial t}(\textbf{r}, t),
\end{align}
where $\varphi(t)$ are TD orbitals giving rise to a TD density $\rho(t)$ by substituting them for the GS orbitals in Eq.~\eqref{eq:density}. The Hamiltonian ${\cal H}$ of Eq.~\eqref{eq:ks_td} in the exact formulation of TDDFT~\cite{rung+1984prl} depends on the electron density $\rho$ at all previous times;\cite{mait+2002prl} such memory effects are not taken into account here. Instead, three complementary approximations for ${\cal H}(t)$ are considered:
\begin{subequations}
\begin{align}
    {\cal H}_\mathrm{IPA}(t) &= \textbf{r}\cdot\textbf{E}(t) + {\cal H}_\mathrm{GS}[\rho_\mathrm{GS}](R_\mathrm{GS}) \\
    {\cal H}_\mathrm{AA}(t) &= \textbf{r}\cdot\textbf{E}(t) + {\cal H}_\mathrm{GS}[\rho(t)](R_\mathrm{GS}) \\
    {\cal H}_\mathrm{AA+E}(t) &=   \textbf{r}\cdot\textbf{E}(t)+{\cal H}_\mathrm{GS}[\rho(t)](R(t)),
\end{align}    
\end{subequations}
where $\textbf{E}(t)$ is the external electric field at time $t$, coupled to the system in the dipole approximation. In the independent-particle approximation (IPA), the electron-electron interaction potential $V_\mathrm{HXC}$ is frozen in the GS status, using the GS functional in tandem with the GS density $\rho_\mathrm{GS}$. In the adiabatic approximation (AA), it is dynamically approximated by inserting the instantaneous TD density snapshot $\rho(t)$ into the GS HXC potential. In both approximations, the nuclei are fixed in the GS configuration $R_\mathrm{GS}$. In the final expression, the nuclear positions are allowed to change in time. The nuclear motion is calculated with the Ehrenfest scheme, in which the positions of the nuclei evolve in time according to Newton's equation:
\begin{equation}
    M_n\frac{\text d^2\mathbf{R}_n}{\text dt^2} =  Z_n\textbf{E}(t)-\nabla_{\mathbf{R}_n}V[\rho(t)](R(t)), 
\end{equation}
where $M_n$ is the mass of the $n$-th nucleus; the electrostatic interaction potential reads
\begin{align}
V[\rho](R) = & \frac{1}{2}\sum_{m = 1} ^N\sum_{n \neq m} ^N\frac{Z_mZ_n}{|\mathbf{R}_m - \mathbf{R}_n|} - \sum_{n=1}^N Z_n\int\text d^3r \frac{\rho(\mathbf{r})}{|\mathbf{r} - \mathbf{R}_n|},
\end{align}
featuring both nuclear-nuclear repulsion and electron-nuclear attraction. Thus, the nuclei are propagated in parallel with the electrons.

\subsection{Implementation}
\label{sec:implementation}
The workflow of our simulations can be divided into three steps (Fig.~\ref{fig:workflow}). During pre-processing, a nested folder structure is set up, corresponding to three different stages with one, two, and three pulses, respectively.
A directory is created for each phase $\varphi$, where the calculations featuring only the first pump pulse are performed. Each of these directories has a subdirectory for every coherence time $\tau$, in which the two-pulse simulations are conducted. Finally, the time propagations with the three-pulse train for different waiting times $T$ are carried out in another series of subfolders within these subdirectories. It is computationally convenient to run the simulations sequentially in accordance to these three layers, \textit{i.e.}, await at each step the conclusion of the previous step. Such a strategy is favorable as it opens up the possibility of eliminating redundancy in the calculations by employing a branching technique. 
This exploits the fact that many of the simulations with different time delays between the three pulses yield identical dipole moments \textit{up to a certain point in time}. For example, two three-pulse calculations with the same coherence time but different waiting times produce the same density up to a point in time beyond the action of the two pump pulses. Thus, there is some redundancy that can be eliminated by running the calculation only once until briefly before that instance, then bifurcating it into two separate ones, one for each waiting time.

We take advantage of this circumstance by dividing the entire simulation into three stages. In the first one, only the first pump pulse is included, followed by a long period of free propagation. During this time, the calculation is periodically interrupted and the status of the calculation is saved and copied whenever the second pump pulse for one of the required coherence times $\tau$ sets on. In the second stage, both pump pulses are included. Several calculations, each for a fixed coherence time $\tau$, start from the milestones generated in the first stage, \textit{i.e.}, just before the respective second pulse interacts with the sample. Also in this case, the system is freely propagated after the second pulse. During this period, the calculation is again paused whenever the third (probe) pulse for one of the desired waiting times $T$ would start, copying the status to provide starting points for the final three-pulse calculations that constitute the third stage. The two-pulse calculations without probe furthermore provide the second-order, pump-only-induced dipole moment eventually to be subtracted from the total one. With the stages building upon each other, they have to be conducted sequentially, slightly compromising overall parallelizability. However, within a given stage, the simulations remain independent and, consequently, perfectly parallelizable. Once all simulations are finished, the dipole moments are mixed between the different phases and contaminant components removed by subtraction of pump-only and probe-only polarizations, yielding the $\textbf{k}_3$-emitting part of the third-order dipole moment (Appendix~\ref{sec:appendix_phase_cycling}). Finally, the 2DES $A_\mathrm{2D}(\omega_\mathrm{exc}, T, \omega_\mathrm{det})$ is obtained via Eq.~\eqref{eq:2des_from_dipole}.

In summary, the computational methods described above and integrated into the presented implementation contribute decisively to reducing the numerical costs of the 2DES simulations based on RT-TDDFT. Specifically, branching techniques decrease the cost by about 50\%, depending on the chosen time parameters (Appendix~\ref{sec:branching_example}). On top of this, undersampling, introduced in Sec.~\ref{sec:sampling_and_undersampling}, and the new $2\times$-phase cycling technique, discussed in detail in Appendix~\ref{sec:appendix_phase_cycling}, reduce the numerical efforts by another 50\% each. 
As a final remark, we emphasize the excellent parallelization offered by RT-TDDFT in \textsc{Octopus}~\cite{andr+15pccp} which ensures the scalability of the proposed methods to larger systems than those considered in this work.

\section{Results and Discussion}
\label{sec:results}

\begin{figure*}
    \centering
    \includegraphics[width=0.95\textwidth]
    {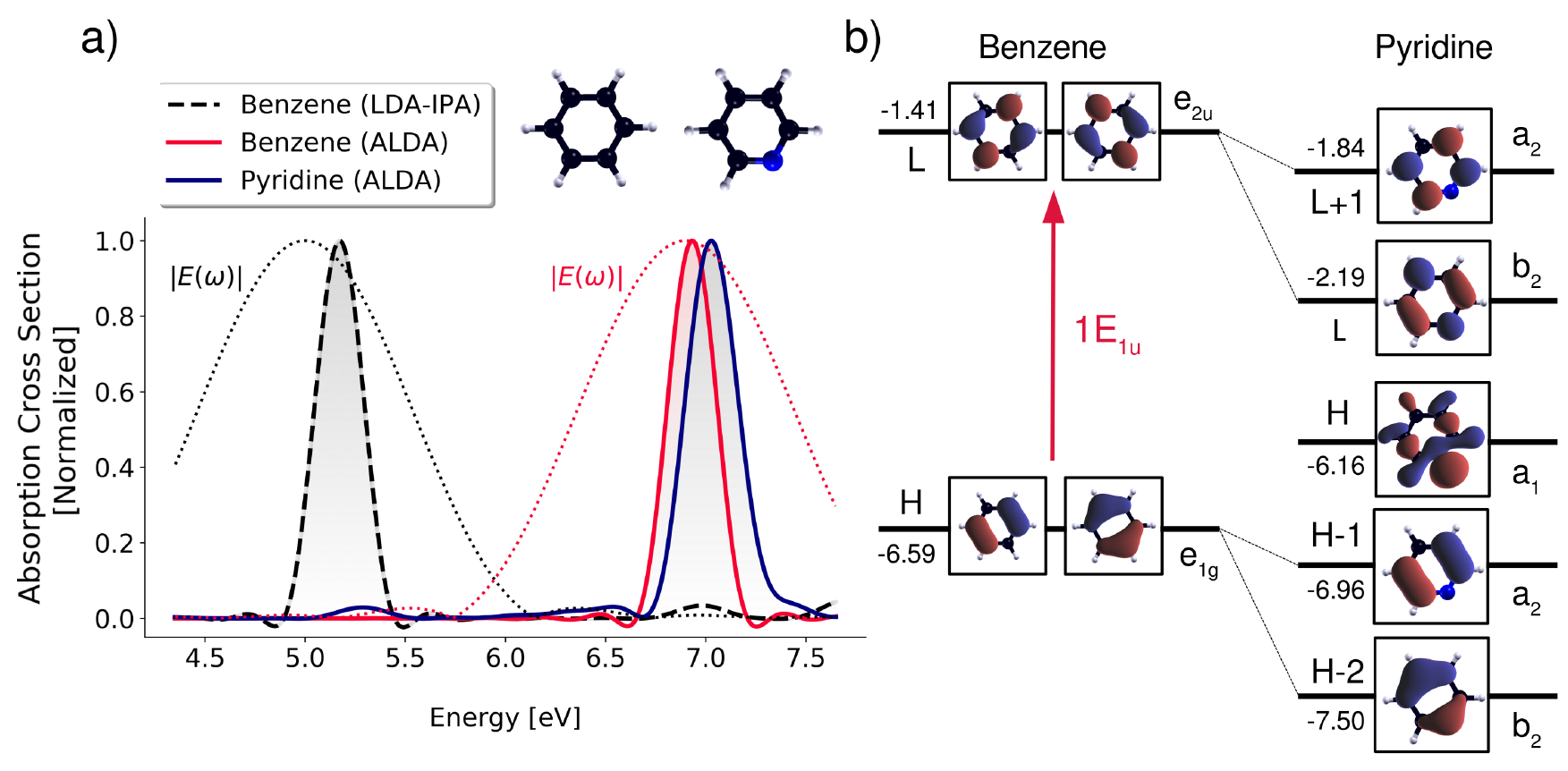}
    \caption{ a) IPA and ALDA linear absorption spectra of benzene (dashed black/solid red) and pyridine (dark blue). In the background, the spectra of the pump lasers employed in the calculation of nonlinear spectra are shown by dotted curves (black for IPA, red for ALDA). b) Single-particle energy levels from Kohn-Sham DFT (LDA functional) with energies in eV, and molecular orbitals of benzene (left) and pyridine (right); the corresponding irreducible representations of the $D_{6h}$ (benzene) and $C_{2v}$ (pyridine) point groups are also given. The isovalue of the Kohn-Sham orbitals is set to $\pm 0.05$ in all cases.
        }
    \label{fig:benzene_linear}
\end{figure*}

In this section, we present calculated 2DES for three test molecules, \textit{i.e.}, benzene, pyridine, and pyrene. Computational details are specified in Appendix~\ref{sec:comp_details}. Benzene is chosen as a well-known aromatic, highly symmetric molecule with an optically dominant excited state,~\cite{doer69jcp,koch+otto1972cpl,phil+81jpb} making it an effective two-level system in the linear regime. Pyridine is the simplest N-doped version of benzene, carrying the same number of electrons while having lower symmetry.~\cite{walk+90cp} On the other hand, pyrene, formed by four fused aromatic rings, presents two excited states with similar optical activity in the UV,~\cite{jone+ashe88jcp} turning it into an effective three-level system in the linear regime. For each compound, we examine the linear absorption characteristics and the frontier orbitals before investigating the third-order spectra. Peculiarities of the adopted approach, \textit{e.g.}, the role of the AA to the dynamical exchange-correlation (XC) potential, are discussed along the way.

\subsection{Benzene and Pyridine}

\subsubsection{Linear Regime}

 Benzene is a hydrogen-saturated six-member carbon ring and has $D_{6h}$ symmetry. Pyridine can be viewed as its N-doped counterpart, with one CH group isoelectronically replaced by an N atom [Fig.~\ref{fig:benzene_linear}a)], reducing the symmetry to $C_{2v}$. Importantly, this point group does not contain the inversion operation, which entails non-zero even-order response functions. 

The symmetry of benzene renders its frontier orbitals degenerate, with the highest occupied (HOMO) and the lowest unoccupied molecular orbital (LUMO) transforming according to the 2D $\mathrm{e_{1g}}$ and $\mathrm{e_{2u}}$ irreducible representations, respectively. The transition from the occupied to the unoccupied frontier orbital gives rise to the very bright and energetically isolated 1E$_\mathrm{1u}\leftarrow\,$1A$_\mathrm{g}$ excitation in the linear regime at 5.2~eV or 6.9~eV, depending on whether dynamical HXC effects are taken into account (Fig.~\ref{fig:benzene_linear}). The result of 6.9~eV, obtained within the adiabatic local-density approximation (ALDA), is in excellent agreement with available measurements.\cite{koch+otto1972cpl} No other features with meaningful oscillator strength are close to this dominant peak. 

In pyridine, the symmetry is broken and the degeneracy of the frontier orbitals lifted [Fig.~\ref{fig:benzene_linear}b)]. As a consequence, the main excitation splits into two, separated by $\sim$0.1~eV, resulting in a broadening of the main absorption band. In addition, there is an occupied orbital in the gap with major contributions from the lone electron pair of the N atom. Consequently, new features are found in the linear absorption spectrum [Fig.~\ref{fig:benzene_linear}a)]. Small peaks associated with the gap state appear at 5.3 and 6.5~eV.

\subsubsection{Nonlinear Regime}

\begin{figure}
    \centering
    \includegraphics[width=0.48\textwidth]
    {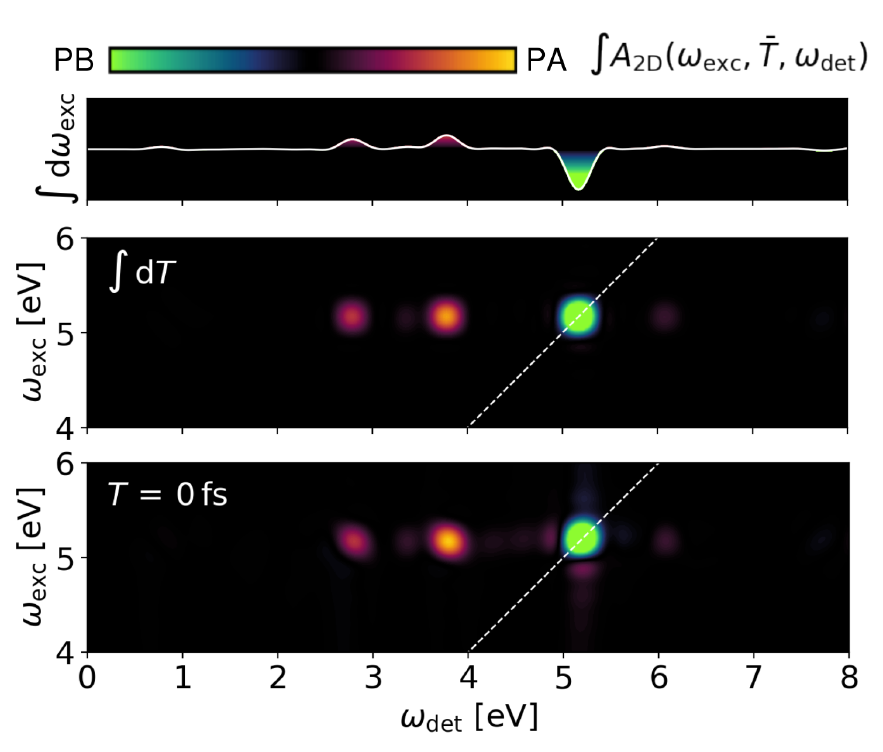}
    \caption{Nonlinear spectra of benzene from LDA-IPA. The upper panel shows the waiting time ($T$) averaged transient-absorption spectrum. It is obtained by integrating over the excitation frequency ($\omega_\mathrm{det}$) the $T$-averaged 2DES, which is displayed in the center panel. The bottom panel shows the $T=0$ snapshot of the 2DES. PA and PB stand for photo-induced absorption and photobleach, respectively.
    }
    \label{fig:2des_benzene_ipa}
\end{figure}

We proceed with the analysis of the excited-state optical properties of benzene and pyridine, starting at the IPA level, \textit{i.e.}, freezing the HXC potential in the GS status. Inspecting the corresponding 2DES (Fig.~\ref{fig:2des_benzene_ipa}), we can single out three dominant resonances along the detection axis for fixed excitation energy $\omega_\mathrm{exc}=5.2$~eV, corresponding to nonlinear processes following the 
HOMO $\rightarrow$ LUMO
transition in benzene. The negative diagonal peak at $\omega_\mathrm{det} = \omega_\mathrm{exc} = 5.2$~eV represents a combination of GSB and SE back to the GS. The two positive features at $\omega_\mathrm{det} = 2.86$~eV and $\omega_\mathrm{det} = 3.78$~eV can be assigned to ESA, specifically to the HOMO $\rightarrow$ HOMO-2 hole transition  and the LUMO $\rightarrow$ LUMO+9 electron transition, respectively. The peaks have a nearly circular shape as a consequence of the raised-cosine damping function $D$ [Eq.~\eqref{eq:damping_function}]; the more common Lorentzian shape would be recovered by instead imposing exponential decay. Comparing the snapshot for fixed waiting time ($T=0$~fs, lower panel) to the $T$ average (center panel), only small differences can be spotted: in the snapshot, faint shadows appear around the main SE/GSB valley. These are coherent artifacts resulting from the overlap between the second pump and the probe pulse, which confuses the time ordering of the three field interactions of optical third-order processes. 

 \begin{figure*}[t]
    \centering
    \includegraphics[width=0.95\textwidth]{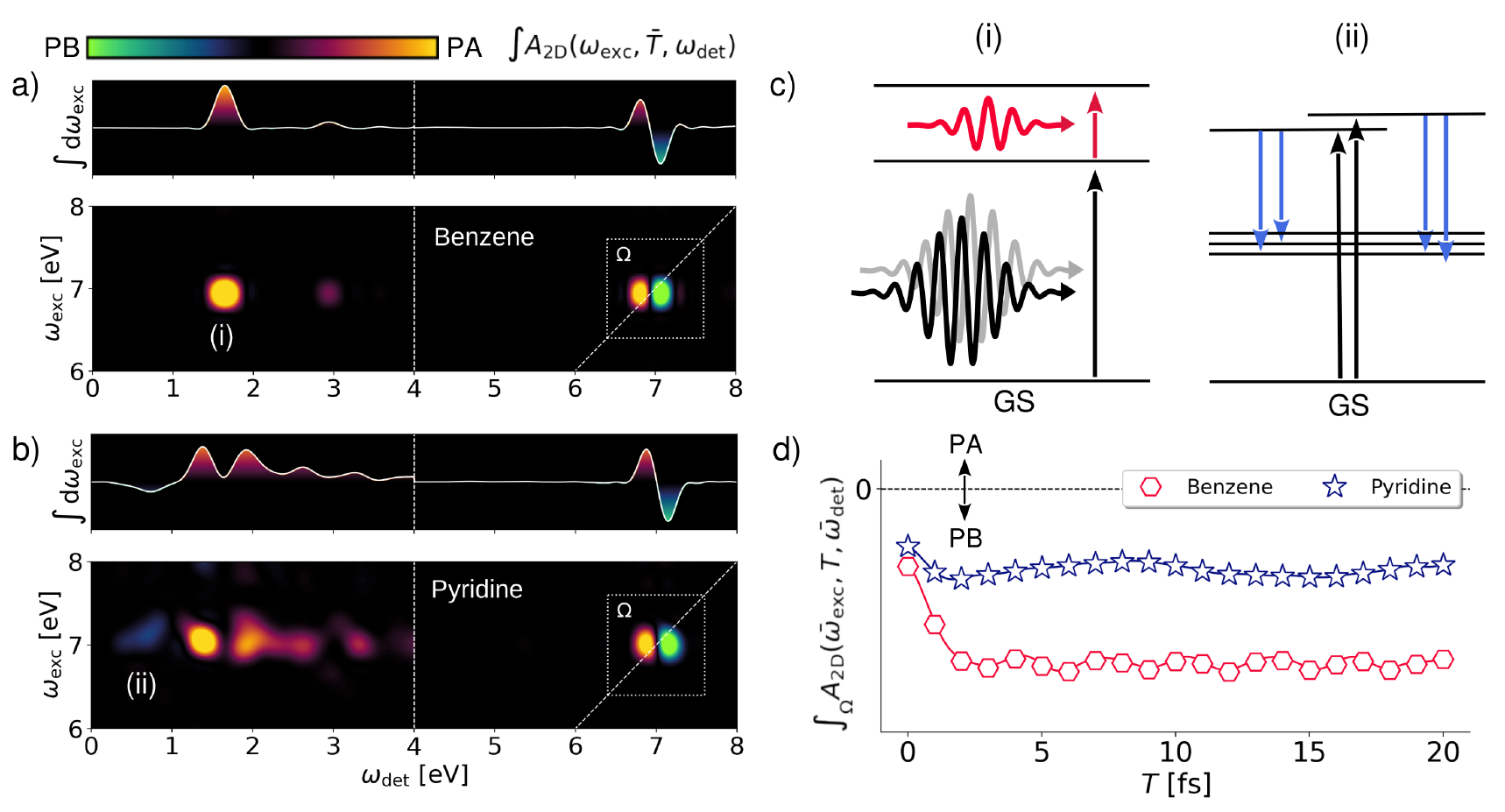}
    \caption{2DES spectra computed from ALDA of a) benzene and b) pyridine averaged over the waiting time $T$ as a function of the excitation frequency $\omega_\mathrm{exc}$ and the detection frequency $\omega_\mathrm{det}$. The maps are magnified by factors of 20 (benzene) and 40 (pyridine) for $\omega_\mathrm{det}<4~\mathrm{eV}$, marked by the vertical dashed line. The dashed squares delimit an integration domain $\Omega$. c) Schemes visualizing
    the ESA and SE processes giving rise to the features labeled (i) and (ii) in panels a) and b), respectively. d) Spectra of benzene and pyridine integrated over the domain $\Omega$ [see panel a)] as a function of waiting time $T$.}
    \label{fig:2des_benzene_alda}
\end{figure*}

Switching on dynamical electron-electron interactions on the ALDA level, the PB feature on the diagonal is blueshifted to 6.9~eV, corresponding to the shift of the linear resonance
[Fig.~\ref{fig:2des_benzene_alda}a)]. It does not exhibit an absorptive lineshape, but more of a dispersive one. This is the result of an artifact originating from the lack of memory of adiabatic XC functionals, which causes a shift of the resonance.\cite{Ruggenthaler_2009, Fuks_2011} When the unshifted linear absorption spectrum is subtracted from the shifted nonlinear one, a derivative lineshape arises. The intensity of such features can be quite large, making it necessary to magnify the low-energy region to visualize the peaks occurring therein. An ESA cross peak is found at $(\omega_\mathrm{det},\,\omega_\mathrm{exc}) = (1.7,\,6.9)$~eV (i), a second faint one at $(3.0,\,6.9)$~eV. The asymmetry in peak height contrasts the IPA case in which the two ESA peaks have comparable heights. This showcases the importance of configuration interaction in the excited states of this molecule. The ultrafast nonlinear optics in the monitored frequency windows are thus essentially determined by only three states [Fig.~\ref{fig:2des_benzene_alda}c(i)]. This is a consequence of a large number of prohibitive selection rules in the highly symmetric molecule.

The low-energy landscape is significantly more diverse in pyridine, giving a first hint about the large size of the optically active space that needs to be considered in nonlinear optics [Fig.~\ref{fig:2des_benzene_alda}b)]. The 0.1~eV split of the main absorption peak is reflected in the spread of the low-energy features along $\omega_\mathrm{exc}$. The nonlinear onset is formed by the negative feature (ii), which relates to SE into three states around 6.3~eV [Fig.~\ref{fig:2des_benzene_alda}c,ii)], which are also bright in the linear regime, albeit with very low absorption strength [Fig.~\ref{fig:benzene_linear}a)].
Meanwhile, the main PB feature on the diagonal at 7~eV again presents the ``peak shift" artifact.
    
As a pragmatic way of dealing with the peak-shift artifact of currently available adiabatic XC functionals, we propose to integrate domains $\Omega$ of the 2DES [outlined in Fig.~\ref{fig:2des_benzene_alda}a)-b)] and plot the result as a function of the waiting time, which is anyway common practice in the analysis of 2DES. If the artifact is a pure shift, it is canceled completely upon integrating the whole feature. Indeed, we find the $\Omega$-integrated values to be strictly negative for both benzene and pyridine [Fig.~\ref{fig:2des_benzene_alda}d)], proving dominant SE and GSB similar to the IPA case. After a transient period of 2~fs during which the probe pulse overlaps the pump pulse, there are only slight modulations in the integrated values. The oscillations are indicative of the participation of other excited states, as will become clear in the case study of the pyrene molecule. However, the small amplitude of these variations confirms that the electronic dynamics are dominated by the respective prevailing bright excitation.

We close this section by mentioning that the ultrafast dynamics of benzene and pyridine have been extensively studied experimentally (see, e.g., Refs.~\citenum{chac-zewa99jpca,yang+20sci} for pyridine and Ref.~\citenum{park+09cpl} for benzene). However, in these works, different excitations were targeted in comparison with those considered in the present study. Specifically, while we pumped the intense resonances around 7~eV in both benzene and pyridine, in the aforementioned references, the weak excitations around 5~eV were stimulated. For this reason, a meaningful comparison between our findings and the experimental data presented in Refs.~\citenum{chac-zewa99jpca,yang+20sci,park+09cpl} is not possible. To the best of our knowledge, experimental 2DES studies of pyridine and benzene are not yet available in the literature. 

\subsection{Pyrene}

\subsubsection{Linear Regime}

\begin{figure*}
    \centering
    \includegraphics[width=0.95\textwidth]{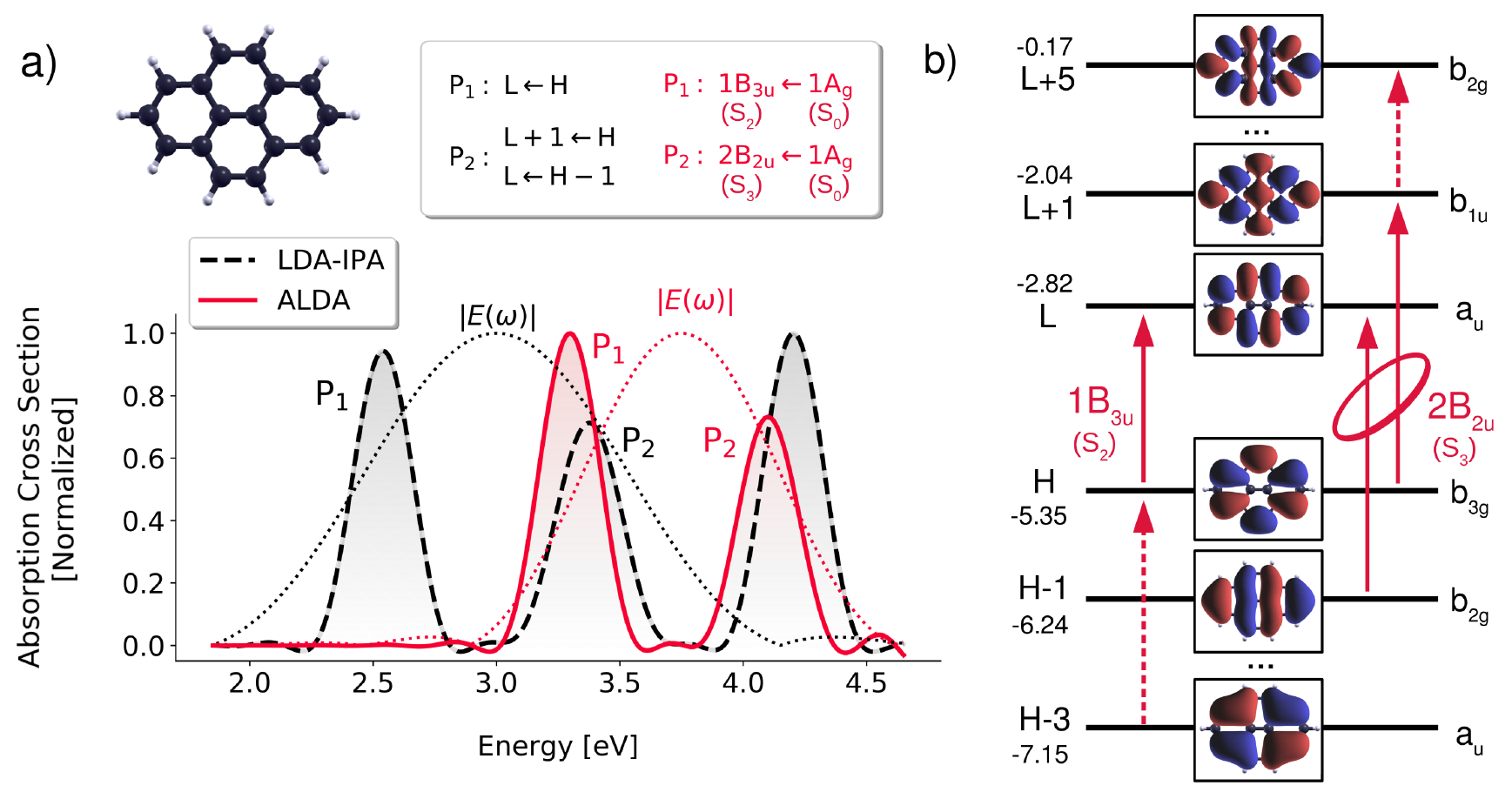}
    \caption{a) Linear absorption spectrum of pyrene (geometry in the top left), at the IPA and ALDA levels of theory. Dotted lines represent the spectra of the pump lasers employed in the calculation of nonlinear spectra. The transitions giving rise to the peaks P$_1$ and P$_2$ are indicated in the inset, using orbital and many-body representations for IPA and ALDA, respectively. b) Molecular orbitals calculated from LDA (energies in eV and irreducible representations of the $D_{2h}$ point group indicated) from pyrene that are involved in the transitions shown in panel a). The solid arrows indicate the single-particle transitions giving rise to P$_1$ and P$_2$; the dashed arrows indicate the dominant excited-state absorption processes. The configurations constituting the excited many-body states 1$\mathrm{B_{3u}}$ (S$_2$) and 2$\mathrm{B_{2u}}$ (S$_3$) are highlighted. The former is of single-determinant character, while the latter features significant configuration interaction, visually represented by the ring. The isovalue of the real-space representations is set to $\pm 0.03$ in all cases.
    }
    \label{fig:pyrene_linear}
\end{figure*}

As a second test system, we consider the pyrene molecule, which has served as a benchmark for 2DES in the ultraviolet spectral range~\cite{Krebs+2013njp} and the theoretical modeling of transient-absorption spectroscopy using quantum-chemical models.\cite{Segatta+2023jctc} This aromatic molecule is constituted by four fused benzene rings and has $D_{2h}$ symmetry (Fig.~\ref{fig:pyrene_linear}). As this point group is Abelian, there are no degenerate states. The orbitals participating in the linear-regime dynamics are the HOMO-1, the HOMO, the LUMO, and the LUMO+1. In the IPA, the lowest energy transition is given by HOMO$\,\rightarrow\,$LUMO at 2.53~eV, giving rise to the peak P$_1$; the energetically close HOMO $\rightarrow$ LUMO+1 (3.31~eV) and HOMO-1 $\rightarrow$ LUMO (3.42~eV) transitions are higher in energy and together constitute peak P$_2$. 

When ALDA interactions are included, the HOMO $\rightarrow$ LUMO transition is blueshifted to 3.3~eV, forming the 1B$_{3u}$ many-body state. Also the two almost resonant HOMO-1 $\rightarrow$ LUMO and HOMO $\rightarrow$ LUMO+1 transitions are, on average, shifted to higher energies. As they have the same symmetry, they are strongly coupled by the electron-electron interaction. Consequently, the resulting 1B$_\mathrm{2u}$ and 2B$_\mathrm{2u}$ many-body states are separated by a large gap of 0.8~eV. The 1B$_\mathrm{2u}\leftarrow\,$1A$_\mathrm{g}$ excitation at 3.3~eV corresponds to the destructive superposition of the constituent single-particle transitions and is effectively dark as a consequence; the 2B$_\mathrm{2u}\leftarrow\,$1A$_\mathrm{g}$ excitation at 4.1~eV corresponds to the constructive superposition and is bright, giving rise to P$_2$. Contrary to the independent-particle case, P$_2$ thus corresponds to a single state in the many-body picture. In the literature, the 1A$_\mathrm{g}$ GS and the 1B$_\mathrm{2u}$, 1B$_\mathrm{3u}$, and 2B$_\mathrm{2u}$ excited states are often referred to as S$_0$, S$_1$, S$_2$, and S$_3$, respectively;~\cite{jone+ashe88jcp} we adopt these labels in the following. The ALDA results for the excitation energies agree within 0.3~eV with those measured in different solvents~\cite{ritt+20acsaem, craw+11jacs, jone+ashe88jcp} and can be improved by employing hybrid XC functionals.\cite{benk+19pccp}
\begin{figure*}
    \centering
    \includegraphics[width=0.95\textwidth]{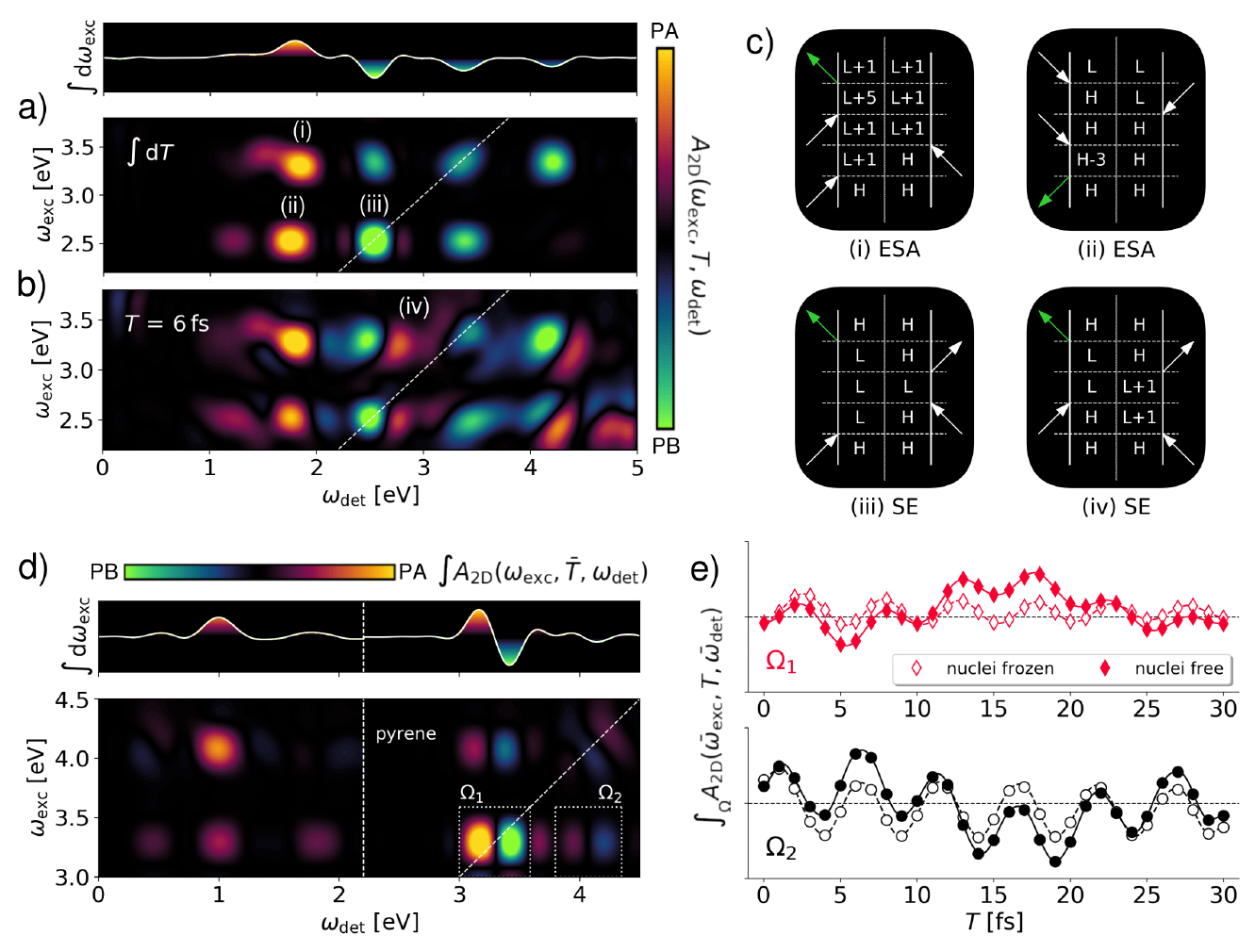}
    \caption{a) 2DES of pyrene calculated from the IPA and integrated over the waiting time $T$. b) Snapshot of the IPA 2DES at the fixed waiting time $T = 6 $~fs. c) Double-sided Feynman diagrams depicting the third-order processes associated with the features in b). The upside-down diagram (ii) represents a hole diagram. d) ALDA 2DES of pyrene integrated over the waiting time $T$ as a function of $\omega_\mathrm{det}$ and $\omega_\mathrm{exc}$. 
    Features below 2.2~eV (marked by the vertical white line) are magnified by a factor of 10. 
    e) 2DES of pyrene computed from ALDA and integrated over the $\Omega_1$ and $\Omega_2$ domains highlighted in a) as a function of $T$. 
    Empty symbols represent exclusively electronic dynamics, whereas the filled ones are the result of RT-TDDFT simulations coupled with Ehrenfest dynamics.}
    \label{fig:pyrene_nonlinear}
\end{figure*}

\subsubsection{Nonlinear Regime}
As before, we first consider the nonlinear dynamics on the IPA level. As this system has multiple bright excitations in the considered frequency band, the 2DES is significantly more complicated than that of the previously investigated benzene (Fig.~\ref{fig:pyrene_nonlinear}). Features appear at the three IPA excitation frequencies corresponding to the HOMO $\rightarrow$ LUMO, HOMO-1 $\rightarrow$ LUMO, and HOMO $\rightarrow$ LUMO+1 transitions, with the latter two having similar, but distinct energies, giving rise to an elliptical peak on the diagonal at $\omega_\mathrm{exc} = \omega_\mathrm{det} = 3.4$~eV [Fig.~\ref{fig:pyrene_nonlinear}a)]. We pick out and analyze in greater detail three of the numerous features in the 2DES, labeled (i)-(iii). (i) is an ESA feature similar to one observed in benzene, where an electron excited from the HOMO to the LUMO+1 is further elevated into the LUMO+5 [Fig.~\ref{fig:pyrene_nonlinear}c)]. (ii) is also ESA, but can be more conveniently modeled as a hole process: the pump pulse pulls a hole from the LUMO into the HOMO, where it can be passed down by the probe pulse into the HOMO-3. (iii) corresponds to overlapping SE and GSB, as usual for diagonal peaks. However, faint positive-valued shadows left and right of the negative peak reveal some superimposed ESA. 

Multiple bright states in the pumped energy window entail rich dynamical features in the 2DES. Inspecting the 2DES for a fixed waiting time, peaks and values appear smeared [Fig.~\ref{fig:pyrene_nonlinear}b)]. This is the result of coherent processes in which the system resides in an excited-state coherence during the waiting time. As an example, we consider the positive feature (iv) in Fig.~\ref{fig:pyrene_nonlinear}b), which is an SE feature. During the waiting time, the system is in a coherent superposition of the HOMO $\rightarrow$ LUMO and HOMO $\rightarrow$ LUMO+1 configurations. A feature resulting from such a pathway will oscillate in time with a period of $1 / (\varepsilon_\mathrm{L+1} - \varepsilon_\mathrm{L}) = 5.3~\mathrm{fs}$, where $\varepsilon_\mathrm{L+1} - \varepsilon_\mathrm{L} = 0.78~\mathrm{eV}$ is the single-particle energy difference between the LUMO+1 and the LUMO. The waiting-time average of such features vanishes, explaining why they cannot be observed in the corresponding spectrum [Fig.~\ref{fig:pyrene_nonlinear}a)].
 
Turning on the dynamical electron-electron interactions, we are once more confronted with the peak-shifting artifact [Fig.~\ref{fig:pyrene_nonlinear}d)], manifesting itself through spurious third-order features along the $\omega_\mathrm{exc}=\omega_\mathrm{det}$ diagonal. It also affects the off-diagonal bleach peaks. The diagonal S$_2$ SE/GSB feature at $\omega_\mathrm{exc}=\omega_\mathrm{det}=3.3$~eV is superimposed with an ESA feature, visible at the high-energy end. Such $\mathrm{S}_n \leftarrow \mathrm{S}_2$ ESA close to the $\mathrm{S}_2 \leftarrow \mathrm{S}_0$ resonance energy has also been predicted with wave function-based methods.~\cite{Segatta+2023jctc} The presence of S$_3$ bleach at $\omega_\mathrm{exc}=\omega_\mathrm{det}=4.1$~eV  is mostly evident from the off-diagonal features, while the diagonal one is very weak. As for the smaller molecules, the ESA peaks at low $\omega_\mathrm{det}$ are magnified in order to make them comparable in strength to the shift features. Standing out is ESA at about $\omega_\mathrm{det}=1.0$~eV for both $\omega_\mathrm{exc}=3.3$~eV (S$_2$) and 4.1~eV (S$_3$). For  $\omega_\mathrm{exc}=3.3$~eV (S$_2$), there is a direct correspondence to the IPA ESA peak, as the target state lying 0.2~eV above S$_3$ has a dominant HOMO-3 $\rightarrow$ HOMO component. For  $\omega_\mathrm{exc}=4.1$~eV (S$_3$), on the other hand, this is not true, and the target state has a quite different configuration with respect to the IPA case. This is the result of the correlated character of the S$_3$ state, for which the single-particle picture loses validity. 

At this point, it is interesting to compare the results in Fig. 7d) to experimental 2DES studies of pyrene. Krebs \textit{et al.} have reported 2DES spectra after S$_2$ excitation.~\cite{Krebs+2013njp} These show, at a waiting time of 1 picosecond, \textit{i.e.}, much beyond the S$_2$ lifetime, signatures of a persistent bleaching of the S$_2$ transition (Fig.~4 in Ref.~\citenum{Krebs+2013njp}), including the observation of vibronic sidebands. Bleaching of this transition is also observed in Fig.~\ref{fig:pyrene_nonlinear}d) of our manuscript [diagonal peak at (3.3, 3.3)~eV]. In addition to the bleaching signatures, the data in Fig.~\ref{fig:pyrene_nonlinear}d), computed at an early time and after resonant excitation of S$_2$ and S$_3$, show different excited state absorption and S$_2$/S$_3$ cross-peaks that have not been observed in experiments yet.

The measurements reported in Ref.~\citenum{Krebs+2013njp} also reveal ESA from the S$_1$ and S$_2$ states at energies well below the S$_2$ resonance. The data in Fig.~\ref{fig:pyrene_nonlinear} show low-energy ESA peaks, too. Since the experimental spectra have been recorded at much longer waiting times, they are likely to be affected by energy relaxation dynamics. As such, a direct comparison to our data is, at this point, challenging. First 2DES maps after S$_3$ excitation have been reported by Picchiotti et al.~\cite{picc+19jpcl} Since energy relaxation processes may affect the spectral lineshape, a direct comparison to Fig.~\ref{fig:pyrene_nonlinear}d) is challenging. To capture such processes, it is mandatory to take nuclear motion into account.

\subsubsection{Vibronic Effects}

To conclude the analysis, we touch briefly upon the effects of nuclear motion on electronic dynamics. To this end, we consider the 2DES of pyrene, integrated over two frequency domains $\Omega_1$ and $\Omega_2$ [Fig.~\ref{fig:pyrene_nonlinear}d)-e)]. $\Omega_1$ encloses features on the diagonal at an energy corresponding to the $\mathrm{S}_2 \leftarrow \mathrm{S}_0$ excitation energy; $\Omega_2$ accommodates off-diagonal features at excitation and detection frequencies pertaining to $\mathrm{S}_2 \leftarrow \mathrm{S}_0$ and $\mathrm{S}_3 \leftarrow \mathrm{S}_0$, respectively. With frozen nuclei, both features oscillate coherently with respect to the waiting time with a period length of about 5~fs [Fig.~\ref{fig:pyrene_nonlinear}e)]. This corresponds to an energy of 0.83~eV, matching the energy difference between peaks $\mathrm{S}_2$ and $\mathrm{S}_3$, which indicates that the oscillation is the result of excited-state coherence between those states. Enabling nuclear motion, the electronically coherent dynamics are superimposed by a slower, likewise periodic modulation. The period length appears to amount to approximately 20~fs, corresponding to a frequency of 0.17~eV or 1680~cm$^{-1}$. Indeed, an $a_g$ C=C stretching mode at this frequency is known to couple to the $\mathrm{S}_2\leftarrow\mathrm{S}_0$ excitation.~\cite{herp+21jpca} A corresponding intensity beating of the SE feature has also been predicted by simulations adopting the quantum-nuclear multi-configurational TD Hartree approach.~\cite{Segatta+2023jctc}
The electronic and vibrational oscillations appear to be independent of each other, likely because the frequencies are quite distinct (0.83~eV vs. 0.17~eV). In principle, however, the Ehrenfest scheme is able to capture some aspects of vibronic coherence, which is a post-Born-Oppenheimer phenomenon characterized by a coupled time evolution of electronically coherent oscillations and wavepacket motion.\cite{rozz+18jpcm}

While the Ehrenfest scheme is easy to use and has the ability to capture some vibrationally coherent phenomena, it is mandatory to keep in mind some of its weaknesses to prevent false conclusions. In the classical limit, the nuclear zero-point energy (ZPE), which is quite large in organic molecules due to the high frequencies of their vibrations, is completely neglected. The ZPE is a key driver of non-adiabatic coupling, enabling transitions between states with different irreducible representations through symmetry-breaking fluctuations. For example, it is known that the populations of both the S$_2$ and the S$_3$ states of pyrene decay substantially within <100~fs due to internal conversion.\cite{neuw+fogg97, borr+18as, roos+18cp, aleo+21jcp, Segatta+2023jctc} Much of the population of both states is transferred to the almost dark S$_1$.
Such internal conversion is not captured by the present single-trajectory Ehrenfest scheme, in which the geometry remains totally symmetric.~\cite{herp+21jpca} However, this effect can be included, to an extent, by adopting a semiclassical approach, by approximating the time evolution of nuclear wavepackets by an ensemble of classical trajectories:\cite{live+21jpcl,krum+22prb} such an extension will be the subject of future work. This approach also introduces vibrational peak broadening, albeit missing some spectroscopic details.~\cite{cres+12tca}
Other aspects of fully quantum-mechanical nuclear dynamics, such as wavepacket splitting, cannot be captured.
This is problematic for laser-induced dynamics, where the wavepacket prepared on an excited-state potential-energy surface should in principle leave the FC region and follow a trajectory almost independent of the remaining GS population. 
Consequently, certain nonlinear spectroscopic features are not correctly reproduced, such as the Stokes shift separating GSB and SE features.
Current developments in the direction of post-Ehrenfest molecular dynamics using coupled trajectories can potentially fix some of these issues in the future.\cite{min+2015prl, gossel+2018jctc}

\section{Summary and conclusions}\label{sec:conclusions}

In summary, we have presented a fully first-principles method to compute 2DES spectra based on RT-TDDFT coupled with Ehrenfest molecular dynamics.
We have presented the methodology and its implementation on top of the \textsc{Octopus} code, illuminating three strategies for reducing the computational workload, \textit{i.e.}, branching, undersampling, and reduced phase cycling. 
We have demonstrated the effectiveness of the framework with the examples of benzene, pyridine, and pyrene. With the aid of these prototypical molecules, we have provided a transparent discussion about the nonlinear regime of excitations that can be optimally unraveled by 2DES.
For pyrene, we have discussed basic effects of electron-vibrational coupling on the spectra, showing the potential of the proposed methodology to deal with multiple degrees of freedom (in this case, electronic and vibrational) and their interplays.
In the discussion, we have explicitly addressed the role of the approximations that enter the choice of the exchange-correlation functional in TDDFT, taking as a ``zero-order'' reference the independent-particle approximation, in which the corresponding interactions are neglected in the dynamics regime. 
Through this comparison, we could analyze in detail the artifacts that may come into play with the specific form of the functional, thus providing a platform for further developments that may allow overcoming them.

We emphasize that the proposed approach hereby applied to small molecules, is general and can be readily employed to more complex systems that are relevant to organic photovoltaics or other technological applications. 
We envision a further improvement of the description of the nuclear part of the dynamical simulations through the inclusion of nuclear quantum effects by adopting an ensemble-based approach.
The extension of the presented formalism to periodic systems is straightforward, given the corresponding infrastructure provided by the \textsc{Octopus} code. Specific issues arising in simulations of extended systems will be addressed in dedicated future studies.

To conclude, we believe that our contribution offers a new tool to simulate 2DES spectra from first principles. 
This option becomes particularly relevant to explore new materials that have not been synthesized and/or measured yet, but also to provide independent references to predict and analyze the outcomes of 2DES experiments.

\section*{Data Availability}

The data supporting this report are provided upon reasonable request.

\begin{acknowledgments}
This work is funded by the German Federal Ministry of Education and Research (Professorinnenprogramm III), the Lower Saxony State (Professorinnen f\"ur Niedersachsen, ``Nieders\"achsisches Vorab – SMART'', and DyNano), and by the German Research Foundation (DFG), project number 182087777 -- CRC 951, 395940726 -- CRC 1372, and 465141364 -- DE 3578/3-1/Li 580/16-1. Computational resources were provided by the North-German Supercomputing Alliance, project bep00076, and by the high-performance computing cluster CARL at the University of Oldenburg, which was funded by the DFG (project number INST 184/157-1 FUGG) and by the Ministry of Science and Culture of the Lower Saxony State.
\end{acknowledgments}

\appendix

\section{List of Abbreviations}
\label{sec:abbreviations}
 \begin{table}[h]
    \centering
    \vspace{0.2cm}
    \begin{tabular}{c|l}
    Abbrev. & Meaning \\
    \hline
        2D & two-dimensional \\
        2DES & two-dimensional electron spectroscopy \\
        AA & adiabatic approximation \\
        (A)LDA & (adiabatic) local-density approximation \\
        DM & density matrix \\
        ESA & excited state absorption \\
        fs & femtosecond(s) \\
        HOMO, H & highest occupied molecular orbital \\
        (H)XC & (Hartree-)exchange-correlation \\
        GS(B) & ground state (bleach) \\
        IPA & independent-particle approximation \\
        KS & Kohn-Sham \\
        LUMO, L & lowest unoccupied molecular orbital \\
        PA & photo-induced absorption \\
        PB & photobleach \\
        (RT-TD)DFT & (real-time time-dependent) density-functional theory \\
        SE & stimulated emission
    \end{tabular}
    \label{tab:my_label}
\end{table}
\section{Phase Cycling}\label{sec:appendix_phase_cycling}
The isolation of the dipole component emitted along the probe direction $\textbf{k}_3$ is accomplished by phase cycling techniques. Each contributing factor to the sum in Eq.~\eqref{eq:fourier_decomposition_polarization} is tagged with an additional factor [Eq.~\eqref{eq:phase_factor}] by adding the constant phase offsets $\varphi_1$, $\varphi_2$, and $\varphi_3$ to the three fields. Here, we consider all polarization components with $|n_1| + |n_2| + |n_3| \leq 3$ and determine how two different phase-cycling schemes affect them. 

In the first one, we run calculations four times, setting $\varphi_3=0$ while cycling $\varphi_1 = \varphi_2 = \varphi$ through the values $(0,\,1,\,2,\,3)\pi/2$; in the second one, we take only the first two, $\varphi=(0,\,1)\pi/2$. In Tables~\ref{tab:n_3_phasecycling} to \ref{tab:n_0_phasecycling}, all terms are specified, including the sum of the phase factors [Eq.~\eqref{eq:phase_factor}] for the two phase-cycling schemes ($4\times$ or $2\times$). If the factor is zero, the term is eliminated by the phase cycling average procedure, \textit{i.e.}, by simply adding the total dipole moment for all sets of phases. Starting with $|n_1| + |n_2| + |n_3| = 3$, we find $4\times$ to be very effective in removing unwanted terms (Table~\ref{tab:n_3_phasecycling}, left). The only remaining undesired terms reflect a three-photon absorption from the probe pulse, which can be removed by subtracting the probe-only dipole. In the $2\times$-case, on the other hand, the cycling procedure itself leaves a multitude of contaminant contributions unaffected (Table~\ref{tab:n_3_phasecycling}, right). However, it turns out that all of them can be eliminated by other means: all terms with more than one probe interaction, $|n_3| > 1$, can be made negligible by choosing the probe intensity much smaller than the pump intensity. In the meanwhile, all terms without probe interaction, $n_3 = 0$, can be removed by subtracting the pump-only dipole moment. We note that while the requirement for the pump-only dipole in this scheme appears to represent an increase in computational costs, it actually comes out as a byproduct upon adopting the branching technique described in Section~\ref{sec:implementation}. The only remaining terms are the desired ones. We thus find that the subtraction of pump-only-induced dipole, the choice of a small probe amplitude, and the phase-cycling elimination complement each other perfectly, isolating the third-order components emitting along $\textbf{k}_3$ already with $2\times$.

Continuing with the lower-order contributions, we turn our attention to $|n_1| + |n_2| + |n_3|= 2$, where $4\times$ leaves up four unwanted terms (Table~\ref{tab:n_2_phasecycling}, left). They vanish automatically if the system has inversion symmetry, as a second-order process constituting a measurable dipole moment features at least one dipole-forbidden transition. If the system does not possess this symmetry, the contaminants can be explicitly removed by subtracting the pump- and probe-only-induced dipoles. This makes $4\times$ universally applicable, in contrast to $2\times$, where numerous terms unaffected by phase cycling cannot be nullified by other means (Table~\ref{tab:n_2_phasecycling}, right). Consequently, $2\times$ is limited to systems with a center of symmetry. As for the remaining two orders, $|n_1| + |n_2| + |n_3|= 1$ (Table~\ref{tab:n_1_phasecycling}) and 0 (Table~\ref{tab:n_0_phasecycling}), the unwanted components can simply be subtracted.

\begin{table*}[]
\raggedright\hspace{4.4cm} $\varphi=(0,\,1,\,2,\,3)\pi/2$ \hspace{4.0cm} $\varphi=(0,\,1)\pi/2$
\vspace{0.2cm}\\
\centering
\begin{tabular}{|p{0.35cm}p{0.35cm}p{0.35cm}|p{1.3cm}|p{1.8cm}|}
\hline
\raggedleft $n_1$ & \raggedleft $n_2$ & \raggedleft $n_3$ & \centering factor & \centering survives? \arraybackslash\\
\hline
\raggedleft -3 & \raggedleft  0 & \raggedleft  0 & \centering $0+0i$ &  \arraybackslash\\
\raggedleft -2 & \raggedleft -1 & \raggedleft  0 & \centering $0+0i$ &  \arraybackslash\\
\raggedleft -2 & \raggedleft  0 & \raggedleft -1 & \centering $0+0i$ &  \arraybackslash\\
\raggedleft -2 & \raggedleft  0 & \raggedleft  1 & \centering $0+0i$ &  \arraybackslash\\
\raggedleft -2 & \raggedleft  1 & \raggedleft  0 & \centering $0+0i$ &  \arraybackslash\\
\raggedleft -1 & \raggedleft -2 & \raggedleft  0 & \centering $0+0i$ &  \arraybackslash\\
\raggedleft -1 & \raggedleft -1 & \raggedleft -1 & \centering $0+0i$ &  \arraybackslash\\
\raggedleft -1 & \raggedleft -1 & \raggedleft  1 & \centering $0+0i$ &  \arraybackslash\\
\raggedleft -1 & \raggedleft  0 & \raggedleft -2 & \centering $0+0i$ &  \arraybackslash\\
\raggedleft -1 & \raggedleft  0 & \raggedleft  2 & \centering $0+0i$ &  \arraybackslash\\
\raggedleft \textbf{-1} & \raggedleft  \textbf{1} & \raggedleft \textbf{-1} & \centering $\textbf{4\,+\,0i}$ & \centering \textbf{yes} \arraybackslash\\
\raggedleft \textbf{-1} & \raggedleft \textbf{1} & \raggedleft \textbf{1} & \centering $\textbf{4\,+\,0i}$ & \centering \textbf{yes} \arraybackslash\\
\raggedleft -1 & \raggedleft  2 & \raggedleft  0 & \centering $0+0i$ &  \arraybackslash\\
\raggedleft  0 & \raggedleft -3 & \raggedleft  0 & \centering $0+0i$ &  \arraybackslash\\
\raggedleft  0 & \raggedleft -2 & \raggedleft -1 & \centering $0+0i$ &  \arraybackslash\\
\raggedleft  0 & \raggedleft -2 & \raggedleft  1 & \centering $0+0i$ &  \arraybackslash\\
\raggedleft  0 & \raggedleft -1 & \raggedleft -2 & \centering $0+0i$ &  \arraybackslash\\
\raggedleft  0 & \raggedleft -1 & \raggedleft  2 & \centering $0+0i$ &  \arraybackslash\\
\raggedleft  0 & \raggedleft  0 & \raggedleft -3 & \centering $4+0i$ & \centering yes, but $\dagger,\star$ \arraybackslash\\
\raggedleft  0 & \raggedleft  0 & \raggedleft  3 & \centering $4+0i$ & \centering yes, but $\dagger,\star$ \arraybackslash\\
\raggedleft  0 & \raggedleft  1 & \raggedleft -2 & \centering $0+0i$ &  \arraybackslash\\
\raggedleft  0 & \raggedleft  1 & \raggedleft  2 & \centering $0+0i$ &  \arraybackslash\\
\raggedleft  0 & \raggedleft  2 & \raggedleft -1 & \centering $0+0i$ &  \arraybackslash\\
\raggedleft  0 & \raggedleft  2 & \raggedleft  1 & \centering $0+0i$ &  \arraybackslash\\
\raggedleft  0 & \raggedleft  3 & \raggedleft  0 & \centering $0+0i$ &  \arraybackslash\\
\raggedleft  1 & \raggedleft -2 & \raggedleft  0 & \centering $0+0i$ &  \arraybackslash\\
\raggedleft \textbf{1} & \raggedleft \textbf{-1} & \raggedleft \textbf{-1} & \centering $\textbf{4\,+\,0i}$ & \centering \textbf{yes} \arraybackslash\\
\raggedleft  \textbf{1} & \raggedleft \textbf{-1} & \raggedleft  \textbf{1} & \centering $\textbf{4\,+\,0i}$ & \centering \textbf{yes} \arraybackslash\\
\raggedleft  1 & \raggedleft  0 & \raggedleft -2 & \centering $0+0i$ &  \arraybackslash\\
\raggedleft  1 & \raggedleft  0 & \raggedleft  2 & \centering $0+0i$ &  \arraybackslash\\
\raggedleft  1 & \raggedleft  1 & \raggedleft -1 & \centering $0+0i$ &  \arraybackslash\\
\raggedleft  1 & \raggedleft  1 & \raggedleft  1 & \centering $0+0i$ &  \arraybackslash\\
\raggedleft  1 & \raggedleft  2 & \raggedleft  0 & \centering $0+0i$ &  \arraybackslash\\
\raggedleft  2 & \raggedleft -1 & \raggedleft  0 & \centering $0+0i$ &  \arraybackslash\\
\raggedleft  2 & \raggedleft  0 & \raggedleft -1 & \centering $0+0i$ &  \arraybackslash\\
\raggedleft  2 & \raggedleft  0 & \raggedleft  1 & \centering $0+0i$ &  \arraybackslash\\
\raggedleft  2 & \raggedleft  1 & \raggedleft  0 & \centering $0+0i$ &  \arraybackslash\\
\raggedleft  3 & \raggedleft  0 & \raggedleft  0 & \centering $0+0i$ &  \arraybackslash\\
\hline
\end{tabular}
\quad\hspace{1cm}
\begin{tabular}{|p{0.35cm}p{0.35cm}p{0.35cm}|p{1.3cm}|p{1.8cm}|}
\hline
\raggedleft $n_1$ & \raggedleft $n_2$ & \raggedleft $n_3$ & \centering factor & \centering survives? \arraybackslash\\
\hline
\raggedleft -3 & \raggedleft  0 & \raggedleft  0 & \centering $1+1i$ & \centering yes, but $\ddagger$ \arraybackslash\\
\raggedleft -2 & \raggedleft -1 & \raggedleft  0 & \centering $1+1i$ & \centering yes, but $\ddagger$ \arraybackslash\\
\raggedleft -2 & \raggedleft  0 & \raggedleft -1 & \centering $0+0i$ &  \arraybackslash \,\\
\raggedleft -2 & \raggedleft  0 & \raggedleft  1 & \centering $0+0i$ &  \arraybackslash \,\\
\raggedleft -2 & \raggedleft  1 & \raggedleft  0 & \centering $1-1i$ & \centering yes, but $\ddagger$ \arraybackslash\\
\raggedleft -1 & \raggedleft -2 & \raggedleft  0 & \centering $1+1i$ & \centering yes, but $\ddagger$ \arraybackslash\\
\raggedleft -1 & \raggedleft -1 & \raggedleft -1 & \centering $0+0i$ &  \arraybackslash\,\\
\raggedleft -1 & \raggedleft -1 & \raggedleft  1 & \centering $0+0i$ &  \arraybackslash\,\\
\raggedleft -1 & \raggedleft  0 & \raggedleft -2 & \centering $1-1i$ & \centering yes, but $\star$ \arraybackslash\\
\raggedleft -1 & \raggedleft  0 & \raggedleft  2 & \centering $1-1i$ & \centering yes, but $\star$ \arraybackslash\\
\raggedleft \textbf{-1} & \raggedleft  \textbf{1} & \raggedleft \textbf{-1} & \centering $\textbf{2\,+\,0i}$ & \centering \textbf{yes} \arraybackslash\\
\raggedleft \textbf{-1} & \raggedleft  \textbf{1} & \raggedleft  \textbf{1} & \centering $\textbf{2\,+\,0i}$ & \centering \textbf{yes} \arraybackslash\\
\raggedleft -1 & \raggedleft  2 & \raggedleft  0 & \centering $1+1i$ & \centering yes, but $\ddagger$ \arraybackslash\\
\raggedleft  0 & \raggedleft -3 & \raggedleft  0 & \centering $1+1i$ & \centering yes, but $\ddagger$ \arraybackslash\\
\raggedleft  0 & \raggedleft -2 & \raggedleft -1 & \centering $0+0i$ &  \arraybackslash\,\\
\raggedleft  0 & \raggedleft -2 & \raggedleft  1 & \centering $0+0i$ &  \arraybackslash\,\\
\raggedleft  0 & \raggedleft -1 & \raggedleft -2 & \centering $1-1i$ & \centering yes, but $\star$ \arraybackslash\\
\raggedleft  0 & \raggedleft -1 & \raggedleft  2 & \centering $1-1i$ & \centering yes, but $\star$ \arraybackslash\\
\raggedleft  0 & \raggedleft  0 & \raggedleft -3 & \centering $2+0i$ & \centering yes, but $\star$ \arraybackslash\\
\raggedleft  0 & \raggedleft  0 & \raggedleft  3 & \centering $2+0i$ & \centering yes, but $\star$ \arraybackslash\\
\raggedleft  0 & \raggedleft  1 & \raggedleft -2 & \centering $1+1i$ & \centering yes, but $\star$ \arraybackslash\\
\raggedleft  0 & \raggedleft  1 & \raggedleft  2 & \centering $1+1i$ & \centering yes, but $\star$ \arraybackslash\\
\raggedleft  0 & \raggedleft  2 & \raggedleft -1 & \centering $0+0i$ &  \arraybackslash\,\\
\raggedleft  0 & \raggedleft  2 & \raggedleft  1 & \centering $0+0i$ &  \arraybackslash\,\\
\raggedleft  0 & \raggedleft  3 & \raggedleft  0 & \centering $1-1i$ & \centering yes, but $\ddagger$ \arraybackslash\\
\raggedleft  1 & \raggedleft -2 & \raggedleft  0 & \centering $1-1i$ & \centering yes, but $\ddagger$ \arraybackslash\\
\raggedleft  \textbf{1} & \raggedleft \textbf{-1} & \raggedleft \textbf{-1} & \centering $\textbf{2\,+\,0i}$ & \centering \textbf{yes} \arraybackslash\\
\raggedleft  \textbf{1} & \raggedleft \textbf{-1} & \raggedleft  \textbf{1} & \centering $\textbf{2\,+\,0i}$ & \centering \textbf{yes} \arraybackslash\\
\raggedleft  1 & \raggedleft  0 & \raggedleft -2 & \centering $1+1i$ & \centering yes, but $\star$ \arraybackslash\\
\raggedleft  1 & \raggedleft  0 & \raggedleft  2 & \centering $1+1i$ & \centering yes, but $\star$ \arraybackslash\\
\raggedleft  1 & \raggedleft  1 & \raggedleft -1 & \centering $0+0i$ &  \arraybackslash\,\\
\raggedleft  1 & \raggedleft  1 & \raggedleft  1 & \centering $0+0i$ &  \arraybackslash\,\\
\raggedleft  1 & \raggedleft  2 & \raggedleft  0 & \centering $1-1i$ & \centering yes, but $\ddagger$ \arraybackslash\\
\raggedleft  2 & \raggedleft -1 & \raggedleft  0 & \centering $1+1i$ & \centering yes, but $\ddagger$ \arraybackslash\\
\raggedleft  2 & \raggedleft  0 & \raggedleft -1 & \centering $0+0i$ &  \arraybackslash\,\\
\raggedleft  2 & \raggedleft  0 & \raggedleft  1 & \centering $0+0i$ &  \arraybackslash\,\\
\raggedleft  2 & \raggedleft  1 & \raggedleft  0 & \centering $1-1i$ & \centering yes, but $\ddagger$ \arraybackslash\\
\raggedleft  3 & \raggedleft  0 & \raggedleft  0 & \centering $1-1i$ & \centering yes, but $\ddagger$ \arraybackslash\\
\hline
\end{tabular}
\caption{Phase factors for polarization Fourier components $|n_1|+|n_2|+|n_3|=3$ upon adoption of 4$\times$ (left) and the proposed 2$\times$phase-cycling schemes (right). All entries with factors $\neq 0$ survive the procedure (default = no). The bold rows reflect the desired components, emitting a field along $\textbf{k}_3$; non-bold survivors are contaminants, but can be removed by other means:
$\star$ adopting a weak probe intensity renders terms with $|n_3|>1$ negligible; $\ddagger$ subtracting the dipole moment from two-pulse, pump-only simulations removes all terms with $n_3=0$; $\dagger$ subtracting the dipole moment from probe-only simulations removes terms with $n_1=n_2=0$.
}
    \label{tab:n_3_phasecycling}
\end{table*}
\begin{table*}[]
\raggedright\hspace{4.4cm} $\varphi=(0,\,1,\,2,\,3)\pi/2$ \hspace{4.0cm} $\varphi=(0,\,1)\pi/2$
\vspace{0.2cm}\\
\centering
\begin{tabular}{|p{0.35cm}p{0.35cm}p{0.35cm}|p{1.3cm}|p{1.8cm}|}
\hline
$\raggedleft n_1$ & $\raggedleft n_2$ & $\raggedleft n_3$ & \centering factor & \centering survives? \arraybackslash\\
\hline
\raggedleft -2 & \raggedleft  0 & \raggedleft  0 & \centering $0+0i$ & \, \arraybackslash\\
\raggedleft -1 & \raggedleft -1 & \raggedleft  0 & \centering $0+0i$ & \, \arraybackslash\\
\raggedleft -1 & \raggedleft  0 & \raggedleft -1 & \centering $0+0i$ & \, \arraybackslash\\
\raggedleft -1 & \raggedleft  0 & \raggedleft  1 & \centering $0+0i$ & \, \arraybackslash\\
\raggedleft -1 & \raggedleft  1 & \raggedleft  0 & \centering $4+0i$ & \centering yes, but $\ddagger,\diamond$ \arraybackslash\\
\raggedleft  0 & \raggedleft -2 & \raggedleft  0 & \centering $0+0i$ & \, \arraybackslash\\
\raggedleft  0 & \raggedleft -1 & \raggedleft -1 & \centering $0+0i$ & \, \arraybackslash\\
\raggedleft  0 & \raggedleft -1 & \raggedleft  1 & \centering $0+0i$ & \, \arraybackslash\\
\raggedleft  0 & \raggedleft  0 & \raggedleft -2 & \centering $4+0i$ & \centering yes, but $\dagger,\star,\diamond$\arraybackslash\\
\raggedleft  0 & \raggedleft  0 & \raggedleft  2 & \centering $4+0i$ & \centering yes, but $\dagger,\star,\diamond$ \arraybackslash\\
\raggedleft  0 & \raggedleft  1 & \raggedleft -1 & \centering $0+0i$ & \, \arraybackslash\\
\raggedleft  0 & \raggedleft  1 & \raggedleft  1 & \centering $0+0i$ & \, \arraybackslash\\
\raggedleft  0 & \raggedleft  2 & \raggedleft  0 & \centering $0+0i$ & \, \arraybackslash\\
\raggedleft  1 & \raggedleft -1 & \raggedleft  0 & \centering $4+0i$ & \centering yes, but $\ddagger,\diamond$ \arraybackslash\\
\raggedleft  1 & \raggedleft  0 & \raggedleft -1 & \centering $0+0i$ & \, \arraybackslash\\
\raggedleft  1 & \raggedleft  0 & \raggedleft  1 & \centering $0+0i$ & \, \arraybackslash\\
\raggedleft  1 & \raggedleft  1 & \raggedleft  0 & \centering $0+0i$ & \, \arraybackslash\\
\raggedleft  2 & \raggedleft  0 & \raggedleft  0 & \centering $0+0i$ & \, \arraybackslash\\
\hline
\end{tabular}
\quad\hspace{1cm}
\begin{tabular}{|p{0.35cm}p{0.35cm}p{0.35cm}|p{1.3cm}|p{1.85cm}|}
\hline
$\raggedleft n_1$ & $\raggedleft n_2$ & $\raggedleft n_3$ & \centering factor & \centering survives? \arraybackslash\\
\hline
\raggedleft -2 & \raggedleft  0 & \raggedleft  0 & \centering $0+0i$ & \, \arraybackslash\\
\raggedleft -1 & \raggedleft -1 & \raggedleft  0 & \centering $0+0i$ & \, \arraybackslash\\
\raggedleft -1 & \raggedleft  0 & \raggedleft -1 & \centering $1-1i$ & \centering yes, but $\diamond$\arraybackslash\\
\raggedleft -1 & \raggedleft  0 & \raggedleft  1 & \centering $1-1i$ & \centering yes, but $\diamond$ \arraybackslash\\
\raggedleft -1 & \raggedleft  1 & \raggedleft  0 & \centering $2+0i$ & \centering yes, but $\ddagger,\diamond$\arraybackslash\\
\raggedleft  0 & \raggedleft -2 & \raggedleft  0 & \centering $0+0i$ & \, \arraybackslash\\
\raggedleft  0 & \raggedleft -1 & \raggedleft -1 & \centering $1-1i$ & \centering yes, but $\diamond$ \arraybackslash\\
\raggedleft  0 & \raggedleft -1 & \raggedleft  1 & \centering $1-1i$ & \centering yes, but $\diamond$ \arraybackslash\\
\raggedleft  0 & \raggedleft  0 & \raggedleft -2 & \centering $2+0i$ & \centering yes, but $\dagger,\star,\diamond$ \arraybackslash\\
\raggedleft  0 & \raggedleft  0 & \raggedleft  2 & \centering $2+0i$ & \centering yes, but $\dagger,\star,\diamond$ \arraybackslash\\
\raggedleft  0 & \raggedleft  1 & \raggedleft -1 & \centering $1+1i$ & \centering yes, but $\diamond$\arraybackslash\\
\raggedleft  0 & \raggedleft  1 & \raggedleft  1 & \centering $1+1i$ & \centering yes, but $\diamond$\arraybackslash\\
\raggedleft  0 & \raggedleft  2 & \raggedleft  0 & \centering $0+0i$ & \, \arraybackslash\\
\raggedleft  1 & \raggedleft -1 & \raggedleft  0 & \centering $2+0i$ & \centering yes, but $\ddagger,\diamond$ \arraybackslash\\
\raggedleft  1 & \raggedleft  0 & \raggedleft -1 & \centering $1+1i$ & \centering yes, but $\diamond$ \arraybackslash\\
\raggedleft  1 & \raggedleft  0 & \raggedleft  1 & \centering $1+1i$ & \centering yes, but $\diamond$ \arraybackslash\\
\raggedleft  1 & \raggedleft  1 & \raggedleft  0 & \centering $0+0i$ & \, \arraybackslash\\
\raggedleft  2 & \raggedleft  0 & \raggedleft  0 & \centering $0+0i$ & \, \arraybackslash\\
\hline
\end{tabular}
\caption{Phase factors for polarization Fourier components with $|n_1|+|n_2|+|n_3|=2$ upon adoption of 4$\times$ (left) and the proposed 2$\times$phase-cycling schemes (right). All entries with factors $\neq 0$ survive the procedure (default = no). All surviving entries are contaminants, but can be removed by other means:
$\star$ adopting a weak probe intensity renders terms with $|n_3|>1$ negligible; $\ddagger$ subtracting the dipole moment from two-pulse, pump-only simulations removes all terms with $n_3=0$; $\dagger$ subtracting the dipole moment from probe-only simulations removes terms with $n_1=n_2=0$; $\diamond$ these terms vanish if the system has inversion symmetry. In fact, all second-order contributions do.
}
    \label{tab:n_2_phasecycling}
\end{table*}
\begin{table*}[]
\raggedright\hspace{4.4cm} $\varphi=(0,\,1,\,2,\,3)\pi/2$ \hspace{4.0cm} $\varphi=(0,\,1)\pi/2$
\vspace{0.2cm}\\
\centering
\begin{tabular}{|p{0.35cm}p{0.35cm}p{0.35cm}|p{1.3cm}|p{1.8cm}|}
\hline
$\raggedleft n_1$ & $\raggedleft n_2$ & $\raggedleft n_3$ & \centering factor & \centering survives? \arraybackslash\\
\hline
\raggedleft -1 & \raggedleft  0 & \raggedleft  0 & \centering $0+0i$ & \, \arraybackslash\\
\raggedleft  0 & \raggedleft -1 & \raggedleft  0 & \centering $0+0i$ & \, \arraybackslash\\
\raggedleft  0 & \raggedleft  0 & \raggedleft -1 & \centering $4+0i$ & \centering yes, but $\dagger$\arraybackslash\\
\raggedleft  0 & \raggedleft  0 & \raggedleft  1 & \centering $4+0i$ & \centering yes, but $\dagger$\arraybackslash\\
\raggedleft  0 & \raggedleft  1 & \raggedleft  0 & \centering $0+0i$ & \, \arraybackslash\\
\raggedleft  1 & \raggedleft  0 & \raggedleft  0 & \centering $0+0i$ & \, \arraybackslash\\
\hline
\end{tabular}
\quad\hspace{1cm}
\begin{tabular}{|p{0.35cm}p{0.35cm}p{0.35cm}|p{1.3cm}|p{1.8cm}|}
\hline
$\raggedleft n_1$ & $\raggedleft n_2$ & $\raggedleft n_3$ & \centering factor & \centering survives? \arraybackslash\\
\hline
\raggedleft -1 & \raggedleft  0 & \raggedleft  0 & \centering $1-1i$ & \centering yes, but $\ddagger$ \arraybackslash\\
\raggedleft  0 & \raggedleft -1 & \raggedleft  0 & \centering $1-1i$ & \centering yes, but $\ddagger$\arraybackslash\\
\raggedleft  0 & \raggedleft  0 & \raggedleft -1 & \centering $2+0i$ & \centering yes, but $\dagger$\arraybackslash\\
\raggedleft  0 & \raggedleft  0 & \raggedleft  1 & \centering $2+0i$ & \centering yes, but $\dagger$\arraybackslash\\
\raggedleft  0 & \raggedleft  1 & \raggedleft  0 & \centering $1+1i$ & \centering yes, but $\ddagger$ \arraybackslash\\
\raggedleft  1 & \raggedleft  0 & \raggedleft  0 & \centering $1+1i$ & \centering yes, but $\ddagger$ \arraybackslash\\
\hline
\end{tabular}
\caption{Phase factors for polarization with Fourier components with $|n_1|+|n_2|+|n_3|=1$ upon adoption of 4$\times$ (left) and the proposed 2$\times$phase-cycling schemes (right). All entries with factors $\neq 0$ survive the procedure (default = no). All surviving entries are contaminants, but can be removed by other means:
$\ddagger$ subtracting the dipole moment from two-pulse, pump-only simulations removes all terms with $n_3=0$; $\dagger$ subtracting the dipole moment from probe-only simulations removes terms with $n_1=n_2=0$.
}
    \label{tab:n_1_phasecycling}
\end{table*}
\begin{table*}[]
\raggedright\hspace{4.4cm} $\varphi=(0,\,1,\,2,\,3)\pi/2$ \hspace{4.0cm} $\varphi=(0,\,1)\pi/2$
\vspace{0.2cm}\\
\centering
\begin{tabular}{|p{0.35cm}p{0.35cm}p{0.35cm}|p{1.3cm}|p{1.8cm}|}
\hline
$\raggedleft n_1$ & $\raggedleft n_2$ & $\raggedleft n_3$ & \centering factor & \centering survives? \arraybackslash\\
\hline
\raggedleft  0 & \raggedleft  0 & \raggedleft  0 & \centering $4+0i$ & \centering yes, but $\downarrow$\arraybackslash\\
\hline
\end{tabular}
\quad\hspace{1cm}
\begin{tabular}{|p{0.35cm}p{0.35cm}p{0.35cm}|p{1.3cm}|p{1.8cm}|}
\hline
$\raggedleft n_1$ & $\raggedleft n_2$ & $\raggedleft n_3$ & \centering factor & \centering survives? \arraybackslash\\
\hline
\raggedleft  0 & \raggedleft  0 & \raggedleft  0 & \centering $2+0i$ & \centering yes, but $\downarrow$ \arraybackslash\\
\hline
\end{tabular}
\caption{Phase factor for polarization Fourier components $|n_1|+|n_2|+|n_3|=0$ upon adoption of 4$\times$ (left) and the proposed 2$\times$phase-cycling schemes (right). The sole entry is a contaminant corresponding to the static dipole moment, which can simply be subtracted in the $4\times$ case; as $2\times$ is only applicable in the presence of inversion symmetry, there is no static dipole moment.
}
    \label{tab:n_0_phasecycling}
\end{table*}
\section{Branching Technique}
\label{sec:branching_example}
To obtain a rough estimate of the time saved by the branching approach, we consider a simple example with realistic parameters. The coherence time sampling period is $\Delta\tau$ and the waiting times range from 0 to $T_\mathrm{max}$ in steps of $\Delta T$. For simplicity, pulses are assumed to be $\delta$-shaped, in which case the total propagation time ${\cal T}_{ij}$ for coherence time $\tau=i\Delta \tau$ and waiting time $T=j\Delta T$ is 
\begin{align}
    {\cal T}_{ij} = \tau + T + \tau_d = i\Delta \tau + j\Delta T + \tau_d.
\end{align}
Since the maximum sensible $\tau$ is $\tau_d$, we arrive at a total propagation time of 
\begin{align}
    \sum_{i,j}{\cal T}_{ij} = \frac{1}{2}\left(\frac{\tau_d}{\Delta\tau}+1\right)\left(\frac{T_\mathrm{max}}{\Delta T}+1\right)\left(3\tau_d + T_\mathrm{max}\right).
\end{align}
In the proposed framework, on the other hand, the propagation times ${\cal T}_k$ for the different stages $k$ are
\begin{subequations}
\begin{align}
  {\cal T}_1 &= \tau_d\\
  {\cal T}_2 &= \sum_i T_\mathrm{max} = \left(\frac{\tau_d}{\Delta\tau}+1\right)T_\mathrm{max}\\
  {\cal T}_3 &= \sum_{i,j} \tau_d = \left(\frac{\tau_d}{\Delta\tau}+1\right)\left(\frac{T_\mathrm{max}}{\Delta T}+1\right)\,\tau_d,
\end{align}
\end{subequations}
amounting to
\begin{align}
\sum_k{\cal T}_k = \tau_d+\left(\frac{\tau_d}{\Delta\tau}+1\right)\left[T_\mathrm{max}+\left(\frac{T_\mathrm{max}}{\Delta T}+1\right)\tau_d\right].
\end{align}
Inserting example values of $\tau_d=15$~fs, $\Delta\tau=1$~fs, $T_\mathrm{max}=20$~fs, and $\Delta T=1$~fs yields a total propagation time of 10,920~fs in the direct approach, which is reduced by about 50\% to 5,375~fs with the branching technique.

\section{Computational Details}\label{sec:comp_details}
The electronic structure calculations are conducted with version 9.2 of the \textsc{Octopus} code~\cite{octopus2020}. Wave functions are represented in real space on a grid obtained by uniformly sampling the union of atom-centered spheres of radius 5~\AA{} with a spacing of 0.24~\AA{}. Geometries are optimized with the FIRE algorithm~\cite{fire} until forces are below 10$^{-3}$~eV/\AA{}. The Perdew-Zunger parametrization~\cite{pz} of the local-density approximation (LDA) is adopted to approximate XC effects in GS calculations in conjunction with norm-conserving Troullier-Martins pseudopotentials~\cite{trou-mart91prb}. For time evolutions featuring dynamical electron-electron interactions, its adiabatic TD extension (ALDA) is employed. The propagation is carried out with the approximated enforced time-reversal symmetry scheme~\cite{castro+2004jcp}, using a time step of 2.7 attoseconds. The dephasing time $\tau_d$ is set to 15 fs in all cases. 

The electric fields $\textbf{E}_j$ are enveloped with raised cosines with a temporal width of $\Delta t_j$:
\begin{align}
    \textbf{E}_j(t) = \textbf{E}_{j,0}\theta(\Delta t_j - |t|)\cos^2\left(\frac{\pi t}{2\Delta t_j}\right)\cos(\omega_{j} t + \varphi_j),
\end{align}
where $\textbf{E}_{j,0}$ features the amplitude and the polarization of the pulses. The former is chosen to correspond an intensity of about $10$ GW/cm$^2$ for $j\in\{1,\,2\}$ and ten times as low for $j=3$. $\Delta t_j$ and $\omega_j$ are set to yield laser spectra covering the excitations of interest. All polarizations are set to $(1, 1, 0)/\sqrt{2}$, coupling the laser to in-plane excitations; those polarized out of plane do not contribute to the UV-visible part of the spectrum due to high confinement energy.

Some issues with the AA and the Ehrenfest scheme are pointed out in the results section of this paper. However, it is important to keep in mind also other known pitfalls of the assumed approximations. DFT is a mean-field method that has proven to be effective for weakly correlated systems. In the case of strong correlations, \textit{e.g.}, in the presence of partially filled $d$ shells in metal-organic compounds, advanced approximations for the XC functional or semi-empirical corrections such as the Hubbard $U$ should be considered. TDDFT in the AA with semi-local approximations for the XC potential is furthermore known to struggle with charge-transfer excitations and excitonic effects in extended systems.~\cite{ullr12book, mait16jcp} Also in this case, hybrid XC functionals can be expected to deliver a more accurate description. For solid-state systems, especially those with strong excitonic effects, equation-of-motion methods based on many-body perturbation theory ($GW$ approximation and the Bethe-Salpeter equation) may pose a viable alternative to RT-TDDFT.~\cite{sang+19jpcm, perf+stef23nl}

\newpage


%

\end{document}